\documentclass[twocolumn]{aastex62}

\usepackage{url}
\usepackage{ulem}
\newcommand{\bjdtdb}{\ensuremath{\rm {BJD_{TDB}}}}
\newcommand{\feh}{\ensuremath{\left[{\rm Fe}/{\rm H}\right]}}

\newcommand{\teff}{\ensuremath{T_{\rm eff}}}
\newcommand{\vsini}{\ensuremath{V\ \mathrm{sin}\ i}}
\newcommand{\logg}{\ensuremath{\log{g_*}}}

\newcommand{\msun}{\ensuremath{\,M_\Sun}}
\newcommand{\rsun}{\ensuremath{\,R_\Sun}}
\newcommand{\lsun}{\ensuremath{\,L_\Sun}}
\newcommand{\mj}{\ensuremath{\,M_{\rm J}}}
\newcommand{\rj}{\ensuremath{\,R_{\rm J}}}

\newcommand{\fave}{\langle F \rangle}
\newcommand{\fluxcgs}{10$^9$ erg s$^{-1}$ cm$^{-2}$}

\newcommand{\muarcsec}{$\mu\arcsec$}

\newcommand{\targetA}{TOI\,564}
\newcommand{\targetAb}{\targetA{}\,b}
\newcommand{\ticnumA}{TIC\,1003831}

\newcommand{\targetB}{TOI\,905}
\newcommand{\targetBb}{\targetB{}\,b}
\newcommand{\ticnumB}{TIC\,261867566}

\newcommand{\tess}{\textit{TESS}}
\newcommand{\gaia}{\textit{Gaia}}

\shorttitle{TOI\,564~\MakeLowercase{b} and TOI\,905~\MakeLowercase{b}}
\shortauthors{Davis et al.}

\begin{document}

\title{TOI\,564\,\MakeLowercase{b} and TOI\,905\,\MakeLowercase{b}: Grazing and Fully Transiting Hot Jupiters Discovered by \tess}



\author[0000-0002-5070-8395]{Allen B. Davis}
\affiliation{Department of Astronomy, Yale University, 52 Hillhouse Avenue, New Haven, CT 06511, USA}
\affiliation{National Science Foundation Graduate Research Fellow}

\author[0000-0002-7846-6981]{Songhu Wang}
\affiliation{Department of Astronomy, Yale University, 52 Hillhouse Avenue, New Haven, CT 06511, USA}
\affiliation{\textit{51 Pegasi b} Fellow}

\author{Matias Jones}
\affiliation{European Southern Observatory, Alonso de C\'ordova 3107, Vitacura, Casilla 19001, Santiago, Chile}


\author[0000-0003-3773-5142]{Jason D.\ Eastman}
\affiliation{Center for Astrophysics ${\rm \mid}$ Harvard {\rm \&} Smithsonian, 60 Garden Street, Cambridge, MA 02138, USA}

\author[0000-0002-3164-9086]{Maximilian N.\ G{\"u}nther}
\affiliation{Department of Physics and Kavli Institute for Astrophysics and Space Research, Massachusetts Institute of Technology, Cambridge, MA 02139, USA}
\affiliation{Juan Carlos Torres Fellow}

\author[0000-0002-3481-9052]{Keivan G. Stassun}
\affiliation{Department of Physics and Astronomy, Vanderbilt University, Nashville, TN 37235, USA}
\affiliation{Department of Physics, Fisk University, Nashville, TN 37208, USA}

\author[0000-0003-3216-0626]{Brett C. Addison}
\affil{University of Southern Queensland, Centre for Astrophysics, West Street, Toowoomba, QLD 4350 Australia}


\author[0000-0001-6588-9574]{Karen A.\ Collins}
\affiliation{Center for Astrophysics ${\rm \mid}$ Harvard {\rm \&} Smithsonian, 60 Garden Street, Cambridge, MA 02138, USA}

\author[0000-0002-8964-8377]{Samuel N. Quinn}
\affiliation{Center for Astrophysics ${\rm \mid}$ Harvard {\rm \&} Smithsonian, 60 Garden Street, Cambridge, MA 02138, USA}

\author[0000-0001-9911-7388]{David W. Latham}
\affiliation{Center for Astrophysics ${\rm \mid}$ Harvard {\rm \&} Smithsonian, 60 Garden Street, Cambridge, MA 02138, USA}

\author[0000-0002-0236-775X]{Trifon Trifonov}
\affiliation{Max-Planck-Institut f\"{u}r Astronomie, K\"{o}nigstuhl 17, D-69117 Heidelberg, Germany}

\author{Sahar Shahaf}
\affiliation{School of Physics and Astronomy, Tel-Aviv University, Tel Aviv 69978, Israel}

\author[0000-0002-3569-3391]{Tsevi Mazeh}
\affiliation{School of Physics and Astronomy, Raymond and Beverly Sackler Faculty of Exact Sciences, Tel Aviv University, Tel Aviv 6997801, Israel}

\author[0000-0002-7084-0529]{Stephen R. Kane}
\affiliation{Department of Earth and Planetary Sciences, University of California, Riverside, CA 92521, USA}

\author[0000-0002-0376-6365]{Xian-Yu Wang}
\affiliation{National Astronomical Observatories, Chinese Academy of Sciences, Beijing 100012, People's Republic of China}
\affiliation{University of Chinese Academy of Sciences, Beijing, 100049, People's Republic of China}

\author[0000-0001-5603-6895]{Thiam-Guan Tan}
\affiliation{Perth Exoplanet Survey Telescope, Perth, Western Australia}

\author[0000-0002-2084-0782]{Andrei Tokovinin}
\affiliation{Cerro Tololo Inter-American Observatory, Casilla 603, La Serena, Chile}

\author[0000-0002-0619-7639]{Carl Ziegler}
\affiliation{Dunlap Institute for Astronomy and Astrophysics, University of Toronto, 50 St. George Street, Toronto, Ontario M5S 3H4, Canada}

\author[0000-0003-1001-0707]{Ren\'{e} Tronsgaard}
\affiliation{DTU Space, National Space Institute, Technical University of Denmark, Elektrovej 328, DK-2800 Kgs. Lyngby, Denmark}

\author[0000-0003-3130-2282]{Sarah Millholland}
\affiliation{Department of Astronomy, Yale University, 52 Hillhouse Avenue, New Haven, CT 06511, USA}
\affiliation{National Science Foundation Graduate Research Fellow}

\author{Bryndis Cruz}
\affiliation{Department of Astronomy, Yale University, 52 Hillhouse Avenue, New Haven, CT 06511, USA}


\author{Perry Berlind} 
\affiliation{Center for Astrophysics ${\rm \mid}$ Harvard {\rm \&} Smithsonian, 60 Garden Street, Cambridge, MA 02138, USA}

\author[0000-0002-2830-5661]{Michael L. Calkins} 
\affiliation{Center for Astrophysics ${\rm \mid}$ Harvard {\rm \&} Smithsonian, 60 Garden Street, Cambridge, MA 02138, USA}

\author[0000-0002-9789-5474]{Gilbert A. Esquerdo}
\affiliation{Center for Astrophysics ${\rm \mid}$ Harvard {\rm \&} Smithsonian, 60 Garden Street, Cambridge, MA 02138, USA}

\author[0000-0003-2781-3207]{Kevin I.\ Collins}
\affiliation{George Mason University, 4400 University Drive, Fairfax, VA, 22030 USA}

\author[0000-0003-2239-0567]{Dennis M. Conti}
\affiliation{American Association of Variable Star Observers, 49 Bay State Road, Cambridge, MA 02138, USA}

\author[0000-0002-5674-2404]{Phil Evans}
\affiliation{El Sauce Observatory, Chile}

\author[0000-0003-0828-6368]{Pablo Lewin}
\affiliation{The Maury Lewin Astronomical Observatory, Glendora, California 91741, USA}

\author{Don J. Radford}
\affiliation{Brierfield Observatory, New South Wales, Australia}

\author{Leonardo A. Paredes}
\affiliation{Physics and Astronomy Department, Georgia State University, Atlanta, GA 30302, USA}

\author[0000-0002-9061-2865]{Todd J. Henry}
\affiliation{RECONS Institute, Chambersburg, PA, USA}

\author{Hodari-Sadiki James}
\affiliation{Physics and Astronomy Department, Georgia State University, Atlanta, GA 30302, USA}

\author{Nicholas M. Law}
\affiliation{Department of Physics and Astronomy, The University of North Carolina at Chapel Hill, Chapel Hill, NC 27599, USA}

\author[0000-0003-3654-1602]{Andrew W. Mann}
\affiliation{Department of Physics and Astronomy, The University of North Carolina at Chapel Hill, Chapel Hill, NC 27599, USA}

\author{C\'esar Brice\~no}
\affiliation{Cerro Tololo Inter-American Observatory, Casilla 603, La Serena, Chile}


\author{George R.~Ricker}
\affiliation{Department of Physics and Kavli Institute for Astrophysics and Space Research, Massachusetts Institute of Technology, Cambridge, MA 02139, USA}

\author{Roland Vanderspek}
\affiliation{Department of Physics and Kavli Institute for Astrophysics and Space Research, Massachusetts Institute of Technology, Cambridge, MA 02139, USA}



\author[0000-0002-6892-6948]{Sara Seager}
\affil{Department of Earth, Atmospheric, and Planetary Sciences, Massachusetts Institute of Technology, Cambridge, MA 02139, USA}
\affil{Department of Physics and Kavli Institute for Astrophysics and Space Research, Massachusetts Institute of Technology, Cambridge, MA 02139, USA}
\affil{Department of Aeronautics and Astronautics, Massachusetts Institute of Technology, Cambridge, MA 02139, USA}

\author[0000-0002-4265-047X]{Joshua N.~Winn}
\affiliation{Department of Astrophysical Sciences, Princeton University, 4 Ivy Lane, Princeton, NJ 08544, USA}

\author{Jon M.~Jenkins}
\affiliation{NASA Ames Research Center, Moffett Field, CA 94035, USA}


\author[0000-0002-8781-2743]{Akshata Krishnamurthy}
\affiliation{Department of Physics and Kavli Institute for Astrophysics and Space Research, Massachusetts Institute of Technology, Cambridge, MA 02139, USA}

\author[0000-0002-7030-9519]{Natalie M. Batalha}
\affiliation{University of California, Santa Cruz, CA, USA}

\author[0000-0002-0040-6815]{Jennifer Burt}
\affiliation{Jet Propulsion Laboratory, California Institute of Technology, 4800 Oak Grove Drive, Pasadena, CA 91109, USA}

\author[0000-0001-8020-7121]{Knicole D. Col\'{o}n}
\affiliation{Exoplanets and Stellar Astrophysics Laboratory, Code 667, NASA Goddard Space Flight Center, Greenbelt, MD 20771, USA}

\author[0000-0001-7010-0937]{Scott Dynes}
\affiliation{Department of Physics and Kavli Institute for Astrophysics and Space Research, Massachusetts Institute of Technology, Cambridge, MA 02139, USA}


\author[0000-0003-1963-9616]{Douglas A. Caldwell}
\affiliation{SETI Institute/NASA Ames Research Center}

\author{Robert Morris}
\affiliation{SETI Institute/NASA Ames Research Center}

\author{Christopher E. Henze}
\affiliation{NASA Ames Research Center, Moffett Field, CA 94035, USA}

\author[0000-0003-2221-0861]{Debra A. Fischer}
\affiliation{Department of Astronomy, Yale University, 52 Hillhouse Avenue, New Haven, CT 06511, USA}


\correspondingauthor{Allen B. Davis}
\email{allen.b.davis@yale.edu}

\begin{abstract} 
We report the discovery and confirmation of two new hot Jupiters discovered by the \textit{Transiting Exoplanet Survey Satellite} (\tess): \targetAb\ and \targetBb. The transits of these two planets were initially observed by \tess\ with orbital periods of 1.651 d and 3.739 d, respectively. We conducted follow-up observations of each system from the ground, including photometry in multiple filters, speckle interferometry, and radial velocity measurements. For \targetAb, our global fitting revealed a classical hot Jupiter with a mass of $1.463^{+0.10}_{-0.096}$~\mj\ and a radius of $1.02^{+0.71}_{-0.29}$~\rj. \targetBb\ is a classical hot Jupiter as well, with a mass of $0.667^{+0.042}_{-0.041}$~\mj\ and radius of $1.171^{+0.053}_{-0.051}$~\rj. Both planets orbit Sun-like, moderately bright, mid-G dwarf stars with $V\sim11$. While \targetBb\ fully transits its star, we found that \targetAb\ has a very high transit impact parameter of $0.994^{+0.083}_{-0.049}$, making it one of only $\sim20$ known systems to exhibit a grazing transit and one of the brightest host stars among them. \targetAb\ is therefore one of the most attractive systems to search for additional non-transiting, smaller planets by exploiting the sensitivity of grazing transits to small changes in inclination and transit duration over the time scale of several years.
\end{abstract} 

\keywords{planetary systems, planets and satellites: detection, stars: individual (\ticnumA, \targetA, \ticnumB, \targetB)}

\section{Introduction}
Transiting hot Jupiters are among the best-studied and the most mysterious classes of exoplanets. Despite the discovery, confirmation, and characterization of hundreds of these worlds, questions persist as to their mechanisms of formation and orbital evolution. It is not known, for instance, whether hot Jupiters formed beyond the ice line and migrated inwards \citep{Lin1996}, or whether they formed close to their present-day orbits \citep{Bodenheimer2000, Batygin2016}. Are they connected evolutionarily to warm Jupiters \citep{Huang2016}? What can we infer about the presence of planetary companions to hot Jupiters, which evidence suggests are rare close to the star \citep{Becker2015, Millholland2016, Canas2019a}, but relatively common farther out \citep{Knutson2014}? What can the atmospheres of hot Jupiters, which are best studied through transit and eclipse observations, tell us about their formation scenarios \citep[e.g.,][]{Oberg2011, Sing2016}? 

Our empirical knowledge of hot Jupiters is based on the foundation of our small but growing sample of these worlds (currently numbering $\sim$250). While small, rocky planets are understood to be present, on average, around every star \citep[e.g.,][]{Fressin2013, Petigura2018}, various studies have found that on the order of only $\sim$0.5\% of stars host a hot Jupiter \citep[e.g.,][]{Howard2012, Petigura2018, Zhou2019}, about $\sim$10\% of which will have a geometry resulting in a visible transit. Transiting hot Jupiters are therefore intrinsically rare, and so given the broad and abiding questions surrounding them, there is value in each additional example found, particularly around stars that are amenable to follow-up observations.

The Rossiter-McLaughlin (RM) effect \citep{Holt1893, Schlesinger1910, Rossiter1924, McLaughlin1924} allows measurement of the sky-projected angle $\lambda$ between a planet's orbital plane and its host star's equator \citep[e.g.,][]{Queloz2000, Addison2018}. RM measurements are most sensitive in systems with deep transits of a planet orbiting a bright or rapidly rotating star. By measuring spin-orbit alignments of many systems, we can probe the processes involved in the formation and migration of exoplanets \citep[e.g.,][]{Lin1996,Bodenheimer2000,FordRasio2008,Naoz2011,WuLithwick2011}, in particular hot \citep{CridaBatygin2014,WinnFabrycky2015} and warm Jupiters \citep{Dong2014}, and compact transiting multi-planet systems \citep{Albrecht2013,Wang2018}.

While the Kepler \citep{Borucki2010} and K2 missions \citep{Howell2014} together examined only $\sim$5\% of the sky, the \tess\ mission \citep{Ricker2015} is conducting a survey of $\sim$80\% of the sky, scanning sector-by-sector for transit signals around the brightest and nearest stars. After \tess\ completes its survey, we will have identified nearly all of the most observationally favorable transiting hot Jupiters that will ever be available to astronomers. Therefore, these planets will serve as the best-possible sample for testing the myriad open questions surrounding hot Jupiters.

To date, five new hot Jupiters initially detected by \tess\ have been confirmed: HD 202772A b \citep{Wang2019}, HD 2685 b \citep{Jones2019}, TOI-150 b \citep{Canas2019b, Kossakowski2019}, HD 271181 b \citep{Kossakowski2019}, and TOI-172 b \citep{Rodriguez2019}. Additionally, the HATNet survey \citep{Bakos2004} detected two transiting hot Jupiter candidates in 2010, HATS-P-69 b and HATS-P-70 b, which were later observed by \tess\ leading to their confirmation \citep{Zhou2019}.

A grazing transit is a transit in which only part of the planet's projected disk occults the stellar disk (formally stated: $b + R_{p}/R_{star} > 1$, where $b$ is the impact parameter, and $R_{p}$ and $R_{star}$ are the radii of the planet and star respectively). Such systems are observationally rare; of the more than 3,000 known transiting planets, only about half a percent exhibit a grazing transit at the $1\sigma$ level or higher \citep[NASA Exoplanet Archive\footnote{http://exoplanetarchive.ipac.caltech.edu};][]{Akeson2013}. \tess\ has detected one grazing transiting planet so far: TOI 216 b, a warm giant planet with an outer companion near the 2:1 resonance, orbiting a $V=12.4$ star \citep{Dawson2019}.

Grazing transiting systems present both upsides and downsides for a system's characterization prospects. On the one hand, the planetary radius is more difficult to measure because of the covariance between the planet size and other transit parameters (primarily the impact parameter) compared to a fully transiting system. For this reason, the inferred radius should perhaps be viewed only as a lower limit with high confidence. Furthermore, grazing systems will exhibit lower RM amplitudes because they cover less of the rotating star's surface compared to a fully transiting planet.

On the other hand, grazing transits afford unique opportunities to probe other aspects of the system. \citet{Ribas2008} attempted to exploit the near-grazing transits of the hot Neptune GJ 436 b \citep{Butler2004} to infer perturbations in the orbital inclination caused by interactions with a putative non-transiting outer planet, GJ 436 c, in a 2:1 mean motion resonance. It was later found that the proposed planet was on an unstable orbit \citep{BeanSeifahrt2008, Demory2009} and there was a lack of expected TTV signals in future transits \citep{Alonso2008, Pont2009, Winn2009}. Nevertheless, the underlying methodology is sound. A close-in hot Jupiter, for instance, will experience both a precession in its periastron as well as a precession in its line of nodes when an additional planet is present in the system. These precessions cause impact parameter variations, which, in the case of grazing transits, change both the transit duration and transit depth dramatically. Systems with grazing transits are therefore prime candidates when seeking to detect non-transiting exoplanets \citep[e.g.,][]{MiraldaEscude2002} and even exomoons \citep{Kipping2009, Kipping2010}. 

WASP-34 b \citep{Smalley2011}, a hot sub-Saturn, was the first planet discovered to have a likely grazing transit (with a confidence of 80\%), and its host star remains the brightest known grazing transit host at $V=10.3$. Other notable grazing transiting planets include hot Jupiters such as HAT-P-27 b/WASP-40 b (\citealt{Beky2011}/\citealt{Anderson2011}), WASP-45 b \citep{Anderson2012}, Kepler-434 b \citep{Almenara2015}, Kepler-447 b \citep{LilloBox2015}, K2-31 b \citep{Grziwa2016}, WASP-140 b \citep{Hellier2017}, Qatar-6b \citep{Alsubai2018}, and NGTS-1 b \citep{Bayliss2018}. A pair of sub-Saturns, WASP-67 b \citep{Hellier2012,Bruno2018} and CoRoT-25 b \citep{Almenara2013} are the smallest known grazing transiting planets.

In this work, we report the discovery and confirmation of two new hot Jupiters detected by \tess, each around relatively bright ($V\sim11$) stars: \targetAb\ and \targetBb. \targetA\ is particularly noteworthy as one of the brightest hosts of a grazing transiting planet, making it highly amenable to follow-up observations. Section \ref{sec:data} describes the observations and data reduction methods. Section \ref{sec:stellar} details the stellar parameters for the host stars. Section \ref{sec:global} presents planetary and system parameters from global analyses. Section \ref{sec:discussion} summarizes these results and places them in context.

\section{Observation and Data Reduction}
\label{sec:data}

\subsection{\tess\ Photometry
\label{ss:TESS Photometry}}

\begin{figure}
\vspace{0cm}\hspace{0cm}
\includegraphics[width=\columnwidth]{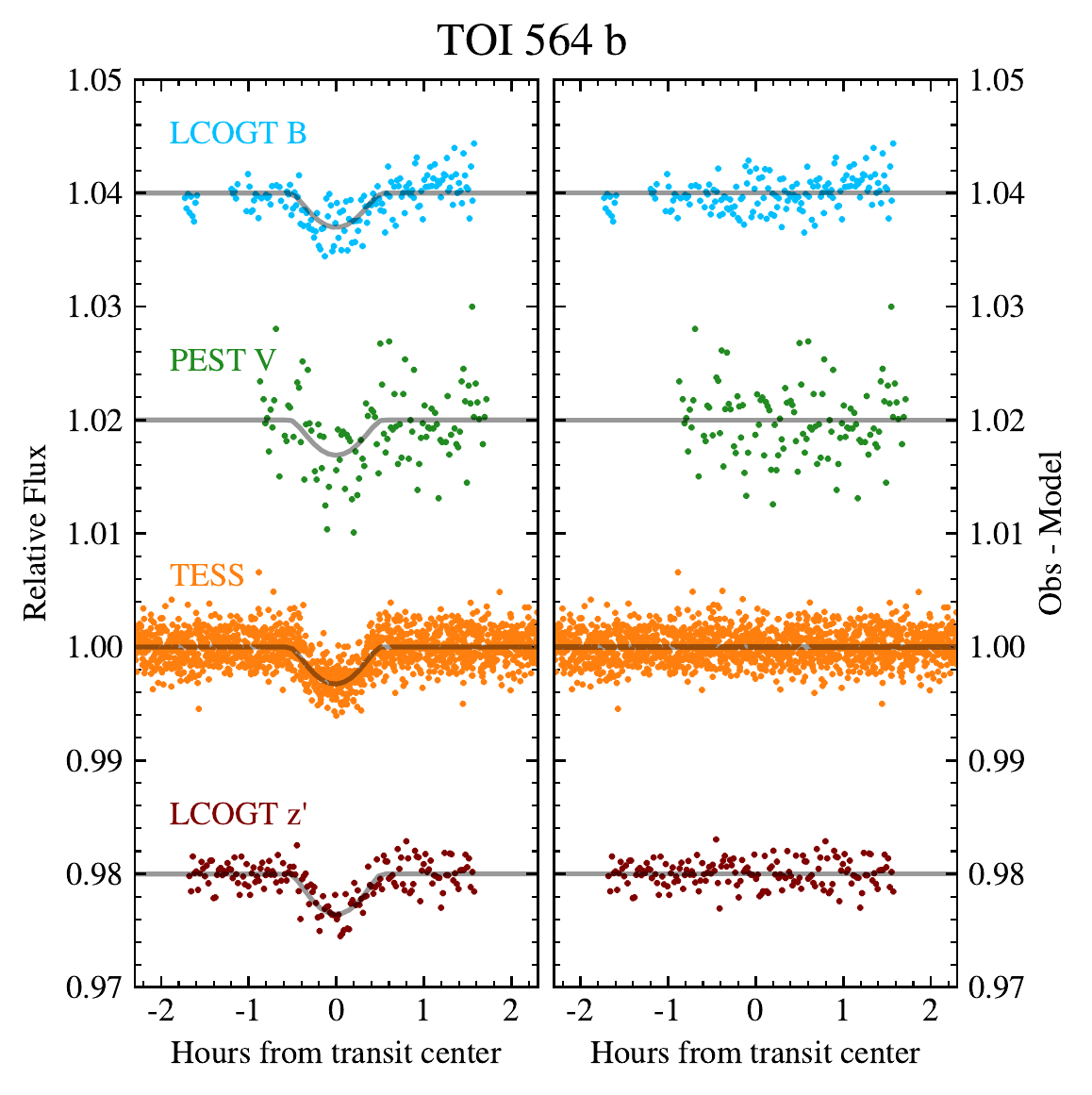}
\caption{Left: Transits of \targetAb. In blue, an LCO-SSO $B$-band light curve from 2019 May 15 (\S \ref{sss:LCOGT}). In green, a PEST $V$-band light curve from 2019 May 10 (\S \ref{sss:PEST}). In orange, the detrended and phase-folded light curve of eleven transits from \tess\ (\S \ref{ss:TESS Photometry}). In maroon, an LCO-CTIO $z'$-band light curve from 2019 April 13 (\S \ref{sss:LCOGT}). The model corresponding to each light curve's filter is shown in grey (\S \ref{sec:global}). Right: Residuals obtained by subtracting the model from the observed transits.
\label{fig:photometry564}}
\end{figure}

\begin{figure}
\vspace{0cm}\hspace{0cm}
\includegraphics[width=\columnwidth]{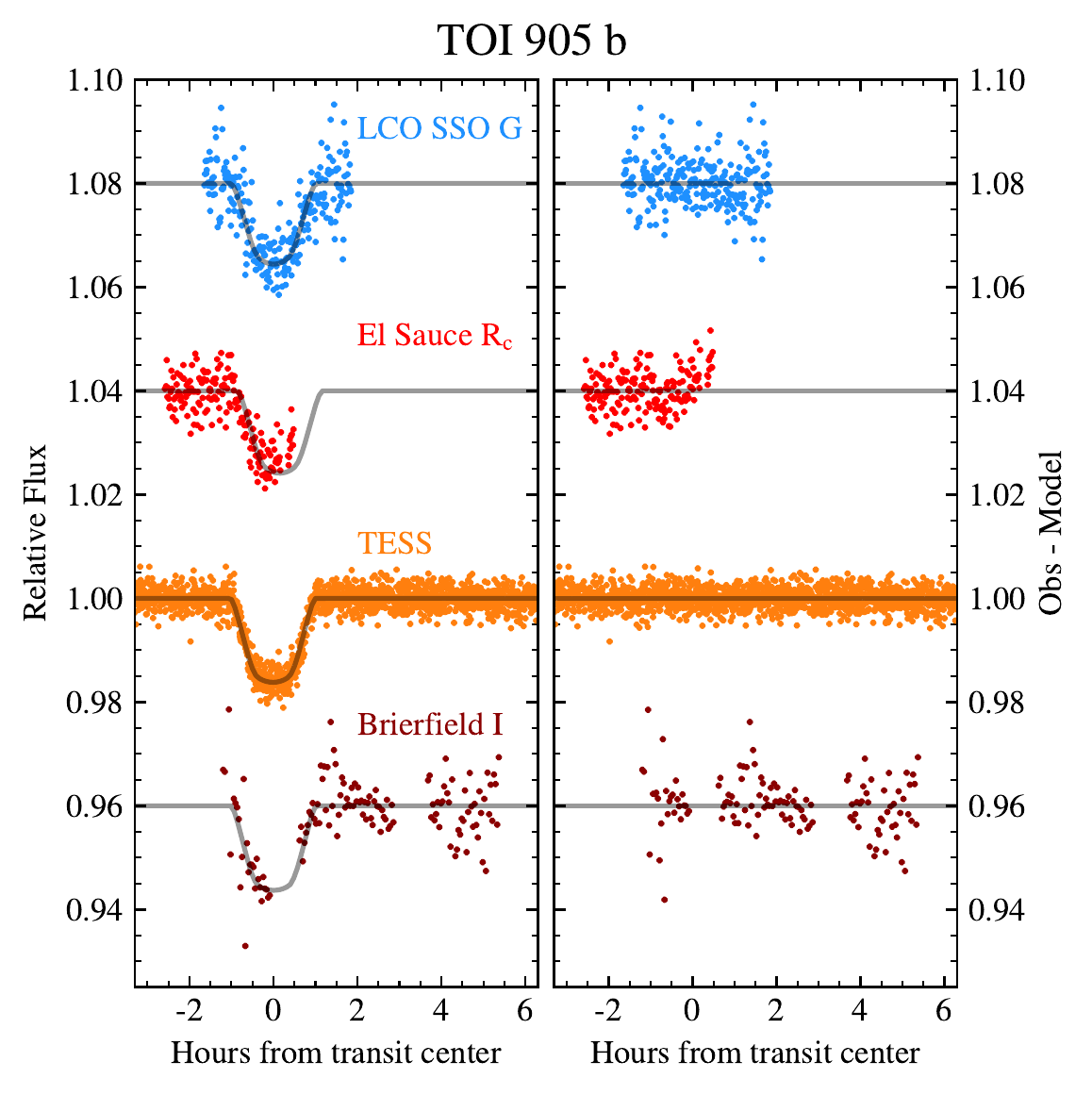}
\caption{Left: Transits of \targetBb. In blue, an LCO-SSO $G$-band light curve from 2019 July 27 (\S \ref{sss:LCOGT}). In red, an El Sauce $R_{\mathrm{c}}$-band light curve from 2019 July 31 (\S \ref{sss:elsauce}). In orange, the detrended and phase-folded light curve of eleven transits from \tess\ (\S \ref{ss:TESS Photometry}). In maroon, a Brierfield $I$-band light curve from 2019 July 27 (\S \ref{sss:brierfield}). The model corresponding to each light curve's filter is shown in grey (\S \ref{sec:global}). Right: Residuals obtained by subtracting the model from the observed transits.
\label{fig:photometry905}}
\end{figure}

\targetA\ (\ticnumA) was observed by \tess\ in Sector 8 by CCD 4 on Camera 2 from 2019 February 2 to 2019 February 27. \targetB\ (\ticnumB) was observed by \tess\ in Sector 12 by CCD 1 on Camera 2 from 2019 May 21 to 2019 June 18. Neither target will be observed again as part of \tess's primary mission. Basic parameters for both targets are given in Table \ref{tab:basic_parameters}.

The photometric data were analyzed with the Science Processing Operations Center (SPOC) pipeline \citep{Jenkins2016} by NASA Ames Research Center. The data have a cadence of two-minutes, and there is a gap of six days in the case of \targetA\ and a gap of one day in the case of \targetB. \tess's CCD pixels have an on-sky size of 21\arcsec. The SPOC pipeline produces two types of light curves: the Simple Aperture Photometry (SAP) light curves, which are corrected for background effects; and the Pre-search Data Conditioning (PDCSAP) light curves \citep{Smith2012, Stumpe2014}, which are additionally corrected for systematics that appear in reference stars.

An automated data validation report \citep[described in ][]{Twicken2018} was created for the PDCSAP light curve of both of our targets, revealing eleven transits on \targetA\ with a period of 1.65114 d and six transits with a period of 3.7395 d for \targetB. This preliminary analysis gave a companion radius of $1.22 \pm 0.16 \rj$ for \targetB\,, consistent with a hot Jupiter. For \targetA\,, the pipeline gave a companion radius of $0.49 \pm 0.24 \rj$, but the impact parameter was extremely poorly constrained; we would later find that this impact parameter was near unity, consistent with a grazing transit, and so our ultimate measurement of the planetary radius was substantially larger than the initial estimate (see Section \ref{sec:global}). Overall, both reports gave highly dispositive results in favor of the planetary hypothesis. The tests used included (for \targetA\ and \targetB\ respectively) the odd-even test ($2.1\sigma$ and $1.6\sigma$ difference), the weak secondary test ($3\sigma$ and $2\sigma$ for the max secondary peak), the statistical bootstrap test (extrapolated FAP $\sim3\times10^{-96}$ and $<10^{-97}$), the ghost diagnostic test (core-to-halo ratio of 3.6 and 5.9), and perhaps most importantly, the difference image centroid offsets from either the TIC position or the out-of-transit centroid (2\arcsec\ in both cases, which is one tenth of a pixel). The difference images are also extremely clean and consistent with the difference image centroids, demonstrating that each the transit source is collocated with the target star image to within the resolution of the survey image.

To remove any stellar variability and other systematics that remained in the PDCSAP light curves, we further detrended the data using the following approach \citep[see, e.g.,][]{Guenther2017, Guenther2018}.
First, we masked out the in-transit data. Then we trained a Gaussian Process (GP) model with a Matern 3/2 kernel and a white noise kernel on the out-of-transit data, using the {\scshape celerite} package \citep{Foreman-Mackey2017}. 
After constraining the hyper-parameters of the GP this way, we applied the GP to detrend the entire light curve. The resulting phase-folded \tess\ light curves near the transits of \targetA\ and \targetB\ are shown in orange in Figure \ref{fig:photometry564} and Figure \ref{fig:photometry905} respectively.

\setlength{\tabcolsep}{20pt}
\begin{deluxetable*}{lrrr}[!]
\hspace{-1in}\tabletypesize{\scriptsize}
\tablecaption{Basic Observational Parameters\label{tab:basic_parameters}}
\tablewidth{0pt}
\tablehead{
\colhead{Parameter}  & \colhead{\targetA} & \colhead{\targetB} & \colhead{Source}
}
\startdata
R.A. (hh:mm:ss)  &  08:41:10.8368   &  15:10:38.0821   & \gaia\ DR2
\\  
Dec. (dd:mm:ss)  & $-$16:02:10.7789 &  $-$71:21:41.8739 & \gaia\ DR2
\\   
$\mu_{\alpha}$ (mas yr$^{-1}$) & $-$2.508 $\pm$ 0.050 
&$-$25.839 $\pm$ 0.033 & \gaia\ DR2
\\
$\mu_{\delta}$ (mas yr$^{-1}$) &  $-$11.025 $\pm$ 0.04 2&  $-$41.150 $\pm$ 0.051
& \gaia\ DR2
\\
Parallax (mas)   &  4.982 $\pm$ 0.031  &  6.274 $\pm$ 0.028  &  \gaia\ DR2
\\
$TESS$ (mag)        &  10.670 $\pm$ 0.006  & 10.572 $\pm$ 0.006 & TIC V8
\\
$B$ (mag)        &  11.946 $\pm$ 0.138  & 12.358 $\pm$ 0.151 & Tycho
\\
$V$ (mag)        &  11.175 $\pm$  0.103  &  11.192 $\pm$ 0.071  & Tycho
\\
$J$ (mag)          &  10.044 $\pm$ 0.030   & 9.890 $\pm$ 0.020 & 2MASS
\\
$H$ (mag)        &  9.710 $\pm$  0.030     &  9.510 $\pm$ 0.020  & 2MASS
\\
$K$ (mag)        &   9.604 $\pm$  0.020   &  9.448 $\pm$ 0.020  & 2MASS
\\
$W1$ (mag)    &   9.562 $\pm$ 0.023    &  9.372 $\pm$ 0.022  & AllWISE
\\
$W2$ (mag)    &   9.598 $\pm$ 0.020    & 9.433 $\pm$ 0.019  & AllWISE
\\
$W3$ (mag)    &   9.587 $\pm$ 0.041    & 9.291 $\pm$ 0.030  & AllWISE
\\
$W4$ (mag)    &         ---           & 9.151 $\pm$ 0.533  & AllWISE
\\
$G$ (mag)         &    11.142\tablenotemark{a}  & 11.081\tablenotemark{a} & \gaia\ DR2
\\
$G_{BP}$ (mag)    &    11.527\tablenotemark{a}  & 11.509\tablenotemark{a} & \gaia\ DR2
\\
$G_{RP}$ (mag)    &    10.622\tablenotemark{a}  & 10.528\tablenotemark{a} & \gaia\ DR2
\enddata
\tablenotetext{a}{For global fitting, we adopted an uncertainty of 0.020 for each \gaia\ magnitude.}
\end{deluxetable*}

\subsection{Ground-Based Transit Photometry
\label{ss:Ground Photometry}}

\setlength{\tabcolsep}{4pt}
\begin{deluxetable*}{lccccccccccc}[!]
\tabletypesize{\scriptsize}
\tablewidth{0pt}
\tablecaption{Ground-Based Transit Photometric Observations\label{tab:ground_phot}}
\tablehead{
\colhead{Telescope}  & \colhead{Camera} & \colhead{Filter}  &  \colhead{Pixel Scale} & \colhead{Est. PSF} & \colhead{Aperture Radius} & \colhead{Date} & \colhead{Duration} & \colhead{\#}  & \colhead{$\sigma$} &  \\
\colhead{} & \colhead{} & \colhead{} & \colhead{(\arcsec)} &  \colhead{(\arcsec)} & \colhead{(pixel)} & \colhead{(UT)} & \colhead{(minutes)} & \colhead{of obs} & \colhead{(ppt)}
}
\startdata 
\smallskip\\\multicolumn{2}{l}{\targetA:}&\smallskip\\
~~~~LCO-CTIO (1 m) & Sinistro &  $z'$  &  0.389  &   1.68  &   15   & 2019 April 13 &  195 & 165 & 0.9
\\  
~~~~MLAO (0.356 m) & STF-8300M &  $B$  &  0.839  &   4.26  &   9   & 2019 May 2 &  82.2 & 48 & 12.0
\\  
~~~~PEST (0.3 m) & ST-8XME &  $V$  &  1.23  &   4.1  &   7   & 2019 May 10 &  155 & 122 & 3.0
\\
~~~~LCO-SSO (1 m) & Sinistro &  $B$  &  0.389  &   2.04  &   14   & 2019 May 15 &  199 & 169 & 1.0
\\  
\smallskip\\\multicolumn{2}{l}{\targetB:}&\smallskip\\
~~~~LCO-SSO (0.4 m) & SBIG 6303 &  $g$  &  0.571  &   9.59 &   15   & 2019 July 27 &  228 & 278 & 2.5
\\ 
~~~~Brierfield (0.36 m) & Moravian 16803 &  $I$  &  1.47  &   4.7  &   6   & 2019 July 27 &  452 & 137 & 1.7
\\
~~~~El Sauce (0.36 m) & STT1603-3 &  $R_c$  &  1.47  &   4.46  &   6   & 2019 July 31 &  184 & 186 & 1.4
\\
\enddata
\end{deluxetable*}

Ground-based photometric follow-up observations were used to confirm both that the transit signals detected by \tess\ were indeed on the correct stars (\ticnumA\ and \ticnumB) and to ensure that the detections were robust in multiple bands. Four distinct transits of \targetA\ were observed between 2019 April 13 and 2019 May 15 in three unique bands from four ground-based telescopes. Two distinct transits of \targetB\ were observed on July 27 and July 31 in three unique bands from three ground-based telescopes. Figure \ref{fig:photometry564} and Figure \ref{fig:photometry905} show the light curves for each transit observed for \targetA\ and \targetB\ respectively.

We used {\tt \tess\ Transit Finder}, which is a customized version of the {\tt Tapir} software package \citep{Jensen2013}, to schedule all of the following photometric time-series follow-up observations. Table \ref{tab:ground_phot} gives a summary of the observations, which are described in detail in the following sections.

\subsubsection{Las Cumbres Observatory Global Telescope}
\label{sss:LCOGT}

We acquired ground-based time-series follow-up photometry of full transits of \targetA\ on 2019 April 13 in $z'$ band and on 2019 May 15 in $B$ band from two Las Cumbres Observatory Global Telescope (LCOGT) 1.0 m telescopes \citep{Brown2013} located at Cerro Tololo Inter-American Observatory (CTIO) and Siding Spring Observatory (SSO) respectively. Additionally, we observed a full transit of \targetB\ on 2019 July 27 using the LCOGT 0.4 m telescope at SSO in $G$ band. All images were calibrated by the standard LCOGT BANZAI pipeline, and the photometric data were extracted using the {\tt AstroImageJ} ({\tt AIJ}) software package \citep{Collins2017}. The two 1.0 m telescopes used for \targetA\ are each equipped with a $4096\times4096$ LCO SINISTRO camera that each have an image scale of $0\farcs389$ px$^{-1}$. The 0.4 m telescope at SSO used an SBIG 6303 camera with an image scale of $0\farcs571$ px$^{-1}$. For \targetA\,, 165 images were acquired during the 195 minute observation in $z'$ band, and 169 images were acquired over the 199 minute observation in $B$ band. For the $G$-band transit of \targetB\,, 278 images were taken over 228 minutes. 

The \targetA\ light curves show clear transit detections using apertures with radii $\sim5\farcs5$. The nearest known \gaia\ DR2 star is $23\arcsec$ from \targetA\ and 7.2 mag fainter. The FWHM of the target and nearby stars are $\sim$ $1\farcs7$ and $\sim$ $2\farcs0$ in $z$ and $B$ bands, respectively, so the follow-up aperture is negligibly contaminated by known nearby \gaia\ DR2 stars. The $z'$- and $B$-band light curves show events having depths consistent with the \tess\ depth within uncertainties.

The \targetB\ light curve also shows a clear transit detection that is consistent with the \tess\ light curve using a photometric aperture with radius of $8\farcs5$. \gaia\ DR2 finds there is a star that is 6.1 mag fainter located $2\farcs2$ away from \ticnumB.

\subsubsection{Maury Lewin Astronomical Observatory}
\label{sss:Lewin}

We observed a transit of \targetAb\ on 2019 May 2 from the Maury Lewin Astronomical Observatory (MLAO), a home observatory located in Glendora, California, USA, using a 0.356 m F10 Schmidt-Cassegrain Celestron C-14 Edge HD telescope with an SBIG STF-8300 detector and a $B$-band filter. The transit was observed at relatively high airmass, ranging from $\sim2$ to $\sim3$, which resulted in low precision ($\sim12.0$ ppt) and a large trend in the time series data. We fitted and removed this airmass trend and found that the transit's depth and shape was generally consistent with the other three ground-based transits, within the large error bars. Because of the lower precision of this transit observation compared to the other $B$-band transit observed by LCO-SSO, we ultimately do not include the MLAO data in the global fitting in Section \ref{sec:global}.

\subsubsection{Perth Exoplanet Survey Telescope}
\label{sss:PEST}

We observed a full transit of \targetAb\ on 2019 May 10 in $V$-band from the 0.3 m Perth Exoplanet Survey Telescope (PEST). PEST is a home observatory located near the city of Perth, Western Australia. The $1530\times1020$ pixel SBIG ST-8XME camera has an image scale of 1$\farcs$23 px$^{-1}$ resulting in a $31\arcmin\times21\arcmin$ field of view. Image reduction and aperture photometry were performed using the C-Munipack program coupled with custom scripts. The light curve has a precision of $\sim3.0$ ppt, which is easily sufficient to verify that the transit depth is consistent with the other light curves.

\subsubsection{Brierfield Observatory}
\label{sss:brierfield}

We observed a transit of \targetB\ on 2019 July 27 in $I$ band using a 0.36 m telescope (PlaneWave CDK14) at the Brierfield Observatory, a home observatory in Brierfield, New South Wales, Australia. The detector was a Moravian 16803 camera, which provided a pixel scale of $1\farcs47$ px$^{-1}$. Seeing conditions were average with some early high cloud limiting pre-ingress time. We observed a continuous transit using 137 images over 452 minutes. The images were reduced and measured as described in \S \ref{sss:LCOGT} with a photometric aperture of $8\farcs8$.

\subsubsection{El-Sauce}
\label{sss:elsauce}

We observed the ingress and  of a transit of \targetB\ on 2019 July 31 in $R_c$ band with a 0.36 m telescope (PlaneWave CDK14) at the El Sauce Observatory, located in Coquimbo Province, Chile. The detector was an SBIG STT1603-1 CCD with a pixel scale of $4\farcs46$ px$^{-1}$. We acquired 186 images over 184 minutes, which were processed with AstroImageJ. Conditions were excellent with no moon or clouds. However, the camera lost its USB connection shortly before egress, so this part of the transit was not captured in this dataset.

\subsection{High Angular-Resolution Observations
\label{sec:angular_res}}

High angular-resolution observations were used to check both systems for close binary companions (including background stars or bound binary companions). We found that both stars have a faint companion nearby, all located within the apertures of the available photometric observations.

\subsubsection{SOAR/HRCam
\label{ss:speckle}}

\targetA\ was observed using speckle interferometry on 2019 May 18 with HRCam \citep{Tokovinin2018, Ziegler2019} in $I$ band on the SOAR 4.1 m telescope. The detector has a pixel scale of 15.75 mas px$^{-1}$, and the angular resolution was 63 mas. We rule out any companions above this limit (e.g., we can rule out a 5.1 mag companion at $>1$\arcsec\ separation).

\begin{figure}
\vspace{0cm}\hspace{0cm}
\includegraphics[width=\columnwidth]{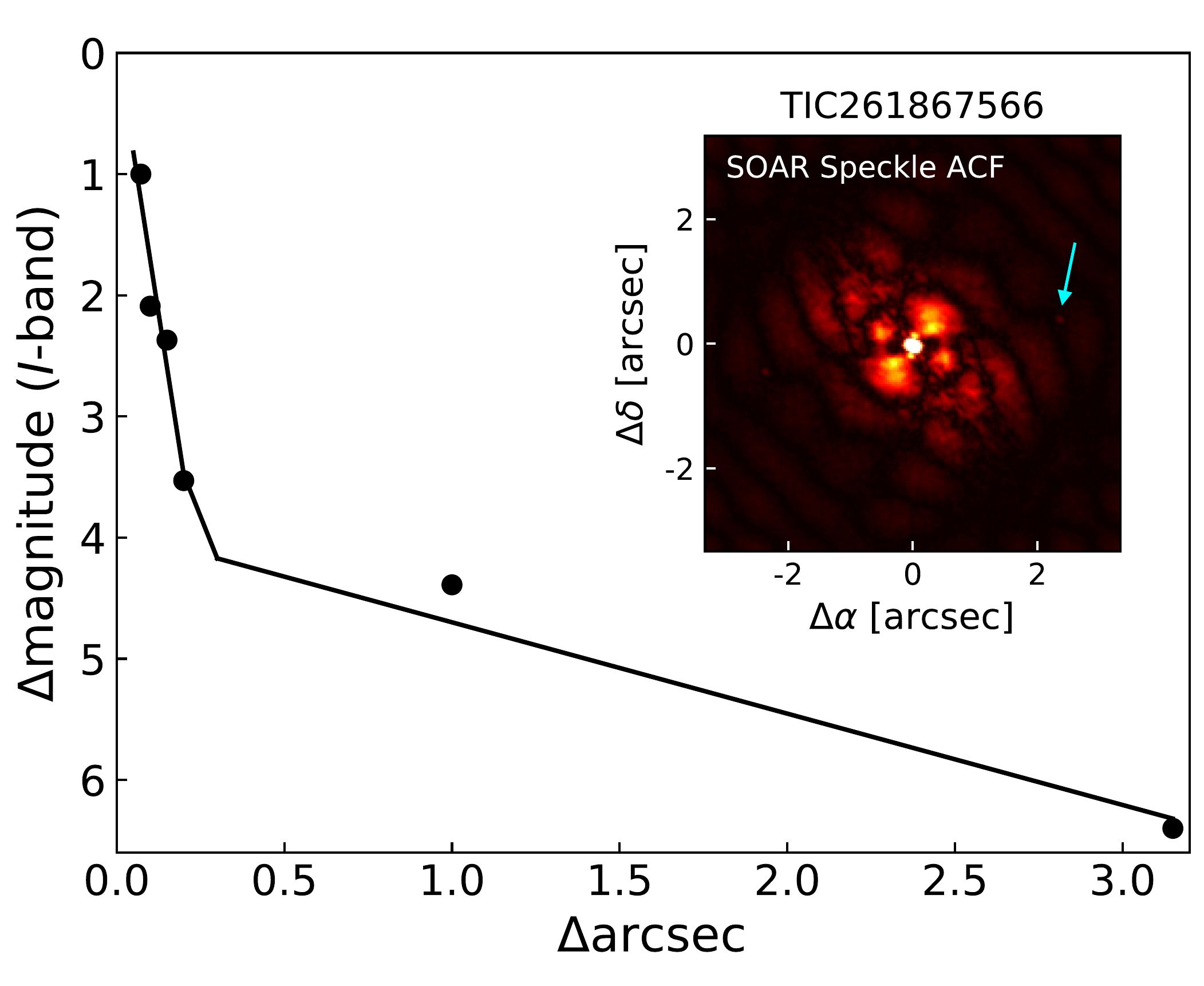}
\caption{HRCam $I$-band contrast curve for \targetB\ with autocorrelation function (ACF) inset. Each point gives the measured $5\sigma$ contrast at various separations from the target, with a smoothing line indicating the expected shape of the contrast curve. The cyan arrow indicates the $\Delta$mag = 5.9 companion $2\farcs28$ away from the primary star.
\label{fig:soar_speckle}}
\end{figure}

HRCam also conducted $I$-band speckle interferometric observations of \targetB\ on 2019 August 12 with an angular resolution of 71 mas. The ACF image in Figure \ref{fig:soar_speckle} shows the 5$\sigma$ detection limit for this target. The HRCam reveals another source located $2\farcs28$ away from \targetB\ that is 5.9 mag fainter in $I$ band. There is no evidence that this companion is physically associated with the system. 


\subsubsection{Palomar 5.1m/PHARO
\label{ss:pharo}}

\begin{figure}
\vspace{0cm}\hspace{0cm}
\includegraphics[width=\columnwidth]{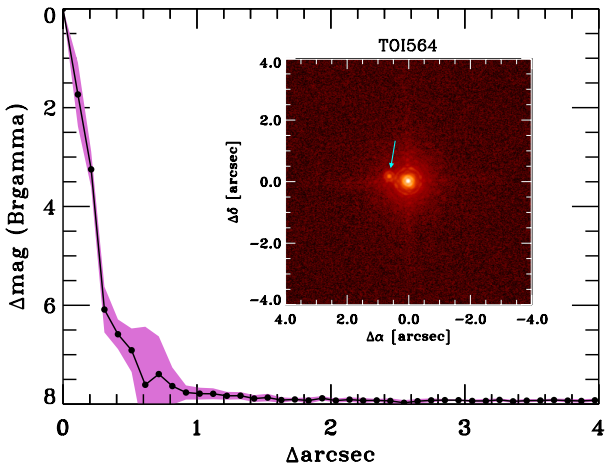}
\caption{PHARO Br-$\gamma$ ($K$ band) contrast curve for \targetA\ with AO image inset. The cyan arrow indicates a $\Delta$mag = 3.53 companion $0\farcs5$ away from the primary star.
\label{fig:pharo}}
\end{figure}

\targetA\ was observed with AO using the Palomar High Angular Resolution Observer (PHARO) on the Palomar-5.1m telescope on 2019 November 10 in $H$ band (continuum) and $K$ band (narrow band Br-$\gamma$). Figure \ref{fig:pharo} reveals a stellar companion is located 0\farcs5 away from the primary star, with $H$ magnitude of 13.40 $\pm$ 0.04 and $K$ magnitude of 13.18 $\pm$ 0.03. These magnitudes and the $H-K$ color are consistent with a early-to-mid M dwarf binary companion with a projected separation of 100 AU, or a giant star 4--5 kpc distant. The former scenario is more parsimonious and has greater potential to create a false positive detection, so we will operate under this assumption.

\subsection{Doppler Measurements
\label{sec:RV}}

\begin{figure*}
\vspace{0cm}\hspace{0cm}
\includegraphics[width=\linewidth]{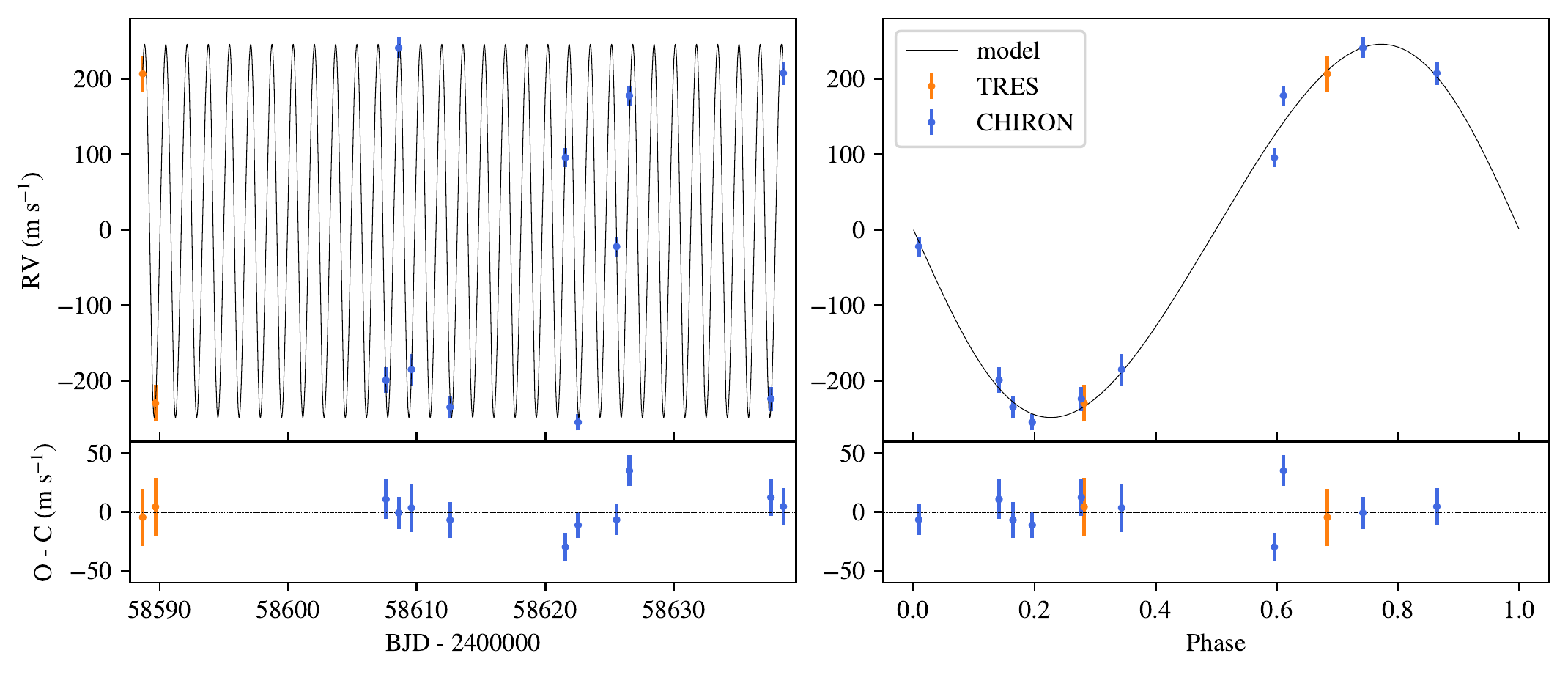}
\caption{Left: Radial velocities of \targetA\ as a function of time, with RVs from TRES and CHIRON plotted in orange and blue respectively. In black, the modeled RV curve based on the median posterior values for parameters derived from the global fitting given in Table \ref{tab:exofast}. TRES RVs were offset to minimize the rms residual from the model determined by CHIRON data. Right: Same as Left, but radial velocity is given as a function of orbital phase. The transit is centered at phase = 0; the closest RV observation to this point did not occur during the transit.
\label{fig:rv_plot564}}
\end{figure*}

\begin{figure*}
\vspace{0cm}\hspace{0cm}
\includegraphics[width=\linewidth]{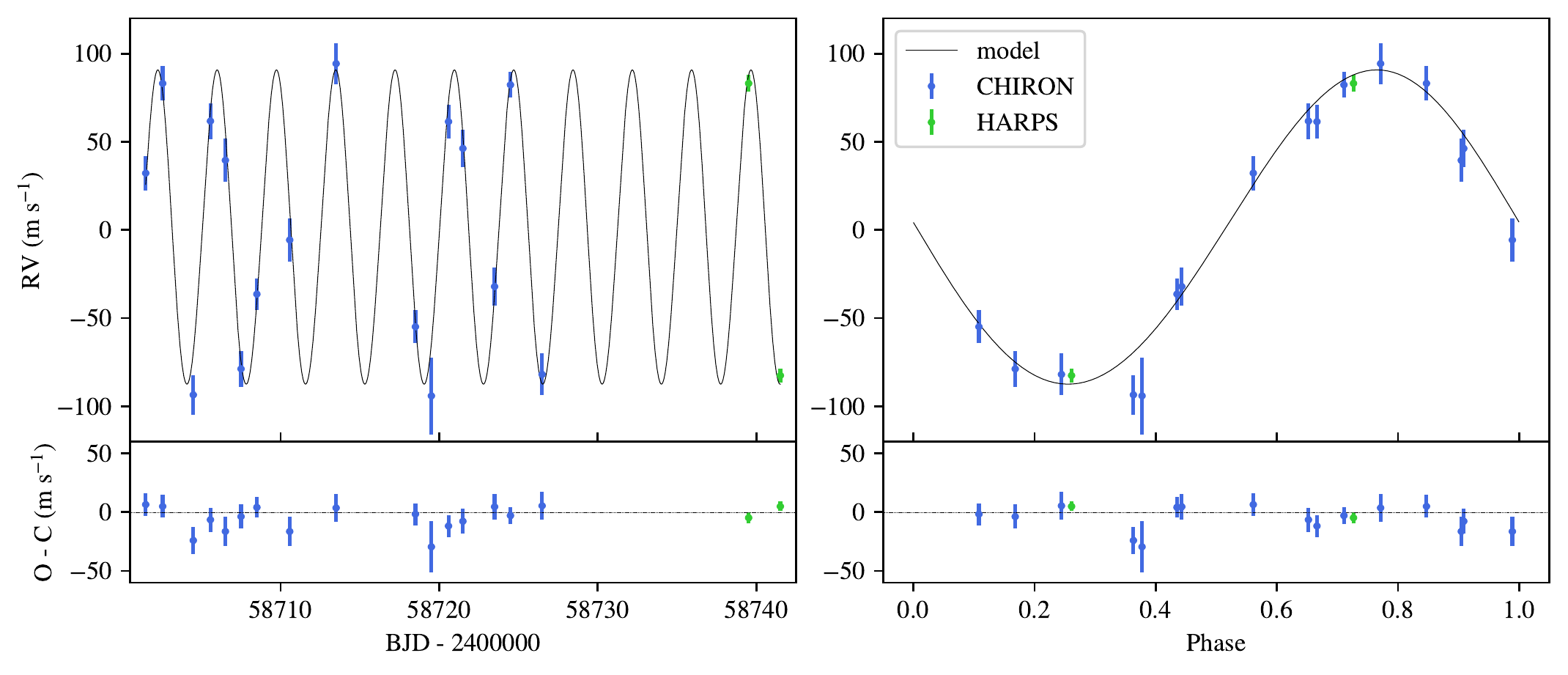}
\caption{Left: Radial velocities of \targetB\ as a function of time, with RVs from CHIRON and HARPS plotted in blue and green respectively. In black, the modeled RV curve based on the median posterior values for parameters derived from the global fitting given in Table \ref{tab:exofast}. HARPS RVs were offset to minimize the rms residual from the model determined by CHIRON data. Right: Same as Left, but radial velocity is given as a function of orbital phase. The transit occurs at phase = 0.
\label{fig:rv_plot905}}
\end{figure*}

We obtained radial velocity (RV) measurements of both systems using three high-precision spectrographs. The velocities for \targetA\ (Figure \ref{fig:rv_plot564}) and \targetB\ (Figure \ref{fig:rv_plot905}) both show strong and clear Keplerian signals, which are discussed in more detail in Section \ref{sec:global}.

\subsubsection{FLWO 1.5m/TRES
\label{ss:TRES_RV}}

We obtained two spectra of \targetA\ with TRES \citep{Furesz2008} on the 1.5 m Tillinghast Reflector telescope at Fred L. Whipple Observatory (FLWO) on Mount Hopkins, AZ, on 2019 April 15 and 16. TRES is an $R \sim 44,000$ echelle spectrograph with a precision of $\sim$10 to 15 m $\mathrm{s^{-1}}$. Spectra are calibrated using a pair of ThAr lamp exposures flanking each set of science exposures. Observations used exposure times of $\sim$ 20 minutes, which yielded a S/N per resolution element of $\sim32$ at 5110 \AA. TRES has an on-sky fiber radius of 1\farcs15.

The reduction and analysis procedures are described in \citet{Buchhave2010}. To summarize, the 2D spectra are optimally extracted and then cross correlated order-by-order using the stronger of the two observations as the template. The radial velocities are determined from a fit to the summed cross-correlation function (CCF), and the internal errors at each epoch are estimated from the standard deviation of the radial velocities derived from the CCF of each order. We also track the instrumental zero point and instrumental precision by monitoring RV standards every night and use this analysis to adjust the RVs and uncertainties. While the internal errors dominate for this star, we do inflate the internal errors by adding the instrumental uncertainty ($\sim$ 10 m $\mathrm{s^{-1}}$) in quadrature. The RVs and uncertainties reported in Table \ref{tab:RV_obs} include these corrections.

\subsubsection{SMARTS 1.5m/CHIRON 
\label{ss:CHIRON}}

\setlength{\tabcolsep}{8pt}
\begin{deluxetable*}{lrrrrrrrll}[!]
\tabletypesize{\scriptsize}
\tablecaption{Radial Velocity Measurements\label{tab:RV_obs}}
\tablehead{
\colhead{BJD\tablenotemark{a} - 2400000}  &  \colhead{RV \tablenotemark{b}}   &   \colhead{$\sigma_\mathrm{{RV}}$}& \colhead{BIS} &  \colhead{$\sigma_\mathrm{{BIS}}$}  &  \colhead{FWHM}  & \colhead{$\sigma_\mathrm{{FWHM}}$} & \colhead{S/N \tablenotemark{c}} & \colhead{Target} & 
\colhead{Instrument}
\\
\colhead{} &\colhead{($\mathrm{m\ s^{-1}}$)} & \colhead{($\mathrm{m\ s^{-1}}$)} & \colhead{($\mathrm{m\ s^{-1}}$)}  &  \colhead{($\mathrm{m\ s^{-1}}$)} & \colhead{($\mathrm{km\ s^{-1}}$)} & \colhead{($\mathrm{km\ s^{-1}}$)} &
\colhead{} &
\colhead{} & \colhead{}
}
\startdata 
58588.6801 &  413.0 &  24.4 & 6.9 &  10.2 & --- & --- & 31.0 & \targetA & TRES
\\
58589.6683 &  -22.8 &  24.4 & -6.9 & 10.2 & --- & --- & 34.2 & \targetA & TRES
\\
58607.5990 & -198.7 &  16.7  &  13.7  &   10.5  &   10.124   &0.153&  37.1 & \targetA & CHIRON
\\  
58608.5907 &  240.8 &  13.6  &   8.6  &    9.1  &   10.441   &0.118&  47.6 & \targetA & CHIRON
\\  
58609.5843 & -184.6 &  20.8  &  27.4  &   12.9  &    9.501   &0.163&  30.9 & \targetA & CHIRON
\\  
58612.5903 & -234.5 &  15.3  & -12.0  &   15.2  &   10.162   &0.129&  38.7 & \targetA & CHIRON
\\  
58621.5585 &  95.6 &  12.3  &  12.0  &   10.2  &   10.088   &0.116&  39.8 & \targetA & CHIRON
\\  
58622.5497 & -254.5 &  10.9  &  13.7  &    9.6  &   10.107   &0.122&  39.4 & \targetA & CHIRON
\\  
58625.5427 &    -22.0 &  13.2  &  30.9  &   10.9  &   10.243   &0.128&  38.4 & \targetA & CHIRON
\\  
58626.5368 &  177.7 &  13.1  &  24.0  &    8.1  &   10.405   &0.125&  48.5 & \targetA & CHIRON
\\  
58637.5439 & -223.5 &  15.9  &  18.9  &   13.5  &   10.241   &0.134&  43.0 & \targetA & CHIRON
\\  
58638.5128 &  207.4 &  15.6  &  22.3  &   15.9  &   10.195   &0.112&  39.9 & \targetA & CHIRON
\\
58701.4994 &  32.2 &  9.7 &  1.3 &  16.5 & 9.953 &  0.073 & 46.8 & \targetB & CHIRON
\\
58702.5685 &  83.1 &   9.7 & -2.6 & 14.1 & 9.885 &  0.079 & 42.2 & \targetB & CHIRON
\\
58704.4989 & -93.5 &   11.3  &   -16.7  &   19.2  &    9.747   &0.100&  36.3 & \targetB & CHIRON
\\  
58705.5794 &  61.7 &  10.3  & -14.1  & 18.9  &  9.864   & 0.094 &  41.1 & \targetB & CHIRON
\\  
58706.5228 & 39.5 &  12.3  &  -7.7  & 22.4 &     9.842  &  0.087 &  38.1 & \targetB & CHIRON
\\  
58707.5088 & -78.8 &  10.2  &  -18.0  & 13.6  & 9.804  & 0.085 &  39.8 & \targetB  & CHIRON
\\  
58708.5073 &  -36.4 &  8.8  & -27.0 &  16.4 &   9.860   &  0.087 &  44.7 & \targetB & CHIRON
\\  
58710.5776 & -5.7 & 12.4 &   0.0  & 20.4  &  9.766   &0.096&  38.8 & \targetB & CHIRON
\\  
58713.5038 &  94.3 &  11.6  &  10.3  &  11.1  &   9.917 &0.088&  44.4 & \targetB & CHIRON
\\  
58718.5033 & -54.9 &   9.4 &  -12.9  &   16.5 &   9.984   & 0.083&  45.8 & \targetB & CHIRON
\\  
58719.5102 & -94.1 &  21.8  & -78.4  &  31.2  &    9.598 &0.108&  28.9 & \targetB & CHIRON
\\  
58720.5899 &  61.4 &  9.4  &  6.4  & 16.8  &   9.951  &0.091&  44.6 & \targetB & CHIRON
\\
58721.4963 & 46.2 &  10.7  &  1.3 & 14.9 &  9.937  & 0.081&  44.1 & \targetB & CHIRON
\\  
58723.4948 & -32.1 &  10.9 & -12.9 & 25.0 &  9.826  & 0.090& 38.9 & \targetB & CHIRON
\\
58724.4959 & 82.3 &  7.3 &  9.0 &  14.4 &  9.968  & 0.087& 45.4 & \targetB & CHIRON
\\
58726.4917 & -81.9 &  11.8  &  38.6 & 20.1 &  9.871 & 0.085 & 40.2 & \targetB & CHIRON
\\
58739.513 & 83.1 &   4.845  &  43.4 & --- &  6.664 & --- & 39.9 & \targetB & HARPS
\\
58741.512 & -82.5 & 4.054 &  15.9 & --- &  6.673 & --- & 44.5 & \targetB & HARPS
\enddata
\tablenotetext{a}{Times are reported according to the BJD at the UTC time at the midpoint of each exposure.}
\tablenotetext{b}{CHIRON RVs are reported with an arbitrary zero point. The zero-point for the TRES and HARPS RVs were each chosen to minimize the least-squared distance from the RV model for the target system based on the global analysis performed on the CHIRON RVs and photometry.}
\tablenotetext{c}{Signal-to-noise ratio per resolution element, reported at 5110 \AA\ for TRES and 5500 \AA\ for CHIRON and HARPS.}
\end{deluxetable*}

We collected ten spectra of \targetA\ with CHIRON \citep{Tokovinin2013}, a fiber-fed spectrograph on the SMARTS 1.5 m telescope at Cerro Tololo, Chile, between 2019 May 4 and 2019 June 4, and sixteen spectra of \targetB\ between 5 August 2019 and 30 August 2019. The short period of both planet candidates allowed us to quickly verify that the star showed an RV signal consistent with a planetary-mass companion by observing each star near the quadrature points implied by the transit ephemerides and the assumption of a circular orbit. We then proceeded to fill out the phase curve of each planet's orbit.

We used CHIRON's $R=80,000$ slicer mode for all observations, which provides substantially higher instrumental throughput when compared to the slit or narrow slit mode \citep[relative efficiencies of the modes are 0.82, 0.25, and 0.11, respectively;][]{Tokovinin2013}. In addition, we did not use the iodine cell, which would have absorbed about half of the stellar light in the $\sim$~5000-6100 \AA\, region. Each observation used an exposure time of 25 minutes, which provided a typical S/N per resolution element of $\sim 40$ at 5500 \AA. The on-sky fiber radius of CHIRON is 1\farcs35.

The RVs were derived closely following the procedure described in \citet{Jones2017} and \citet{Wang2019}. Briefly, we first built a template by stacking the individual CHIRON spectra, after shifting all of them to a common rest frame. We then computed the cross-correlation-function (CCF) between each observed spectrum and the template. The CCF was then fitted with a Gaussian function plus a linear trend. The velocity corresponding to the maximum of the Gaussian fit corresponds to the observed radial velocity. We applied this method to a total of 33 orders, covering the wavelength range of $\sim$ 4700\,-\,6500 \AA. Since CHIRON is not equipped with a simultaneous calibration fiber, we obtained a ThAr lamp immediately before each science observations. The CHIRON pipeline therefore recomputes a new wavelength solution from this calibration observation, thus correcting for the instrumental drift. Using this method, we achieve a long-term stability better than 10 m\,s$^{-1}$, which has been tested using RV standard stars. The final RV at each epoch is obtained from the median in the individual order velocities, after applying a 3$\sigma$ rejection method. The corresponding uncertainty is computed from the error in the mean of the non-rejected velocities (see more details in \citealt{Jones2017}).
The typical RV error found was about 15 $\mathrm{m\ s^{-1}}$. Finally, we also computed the bisector inverse slope (BIS) and full-width-at-half-maximum (FWHM) of the CCF. The full results of the CCF analysis are given in Table \ref{tab:RV_obs}, including the BIS and FWHM diagnostics.

\subsubsection{ESO 3.6m/HARPS
\label{ss:HARPS}}

We collected two spectra of \targetB\ with the High Accuracy Radial velocity Planet Searcher (HARPS) \citep{Mayor2003} at the ESO 3.6 meter telescope. HARPS has a spectral resolution of $R = 115,000$ and a fiber with an on-sky radius of 0\farcs5. Exposure times were 25 minutes, achieving S/N $\sim$ 42 at 5500 \AA.

Our motivation in collecting HARPS spectra was to test if the semiamplitude of the signal was consistent between HARPS and CHIRON, which have sky fibers of 0\farcs5 and 1\farcs35 respectively. If there is potential RV contamination from the nearby star 2\farcs28 away (\S \ref{ss:speckle}), then the Doppler semiamplitude should be different between the instruments. Figure \ref{fig:rv_plot905} shows that when the HARPS RVs are offset to match CHIRON, they agree closely.

\section{Host Star Characterization}
\label{sec:stellar}

It is well-understood that when the transit and RV techniques are used for planet characterization, we can only know the planet as well as we know the star. We derived physical and atmospheric parameters for \targetA\ and \targetB\ using several independent methodologies and data sets that are described in the following subsections (the stellar parameters determined by EXOFASTv2 are described later in Section \ref{sec:global}).

We find that among the values probed by multiple methods there is generally agreement within 1--2$\sigma$, giving us greater confidence in their collective veracity. Both stars are G-type main sequence stars, which are roughly Sun-like in their mass, radius, and temperature. Both stars are metal-rich. A summary of the parameters derived is shown in Table \ref{tab:stellar_params}.

\setlength{\tabcolsep}{10pt}
\begin{deluxetable*}{lcccc}[!]
\tabletypesize{\scriptsize}
\tablecaption{Stellar Parameters\label{tab:stellar_params}}
\tablehead{
\colhead{Parameter}  &  \colhead{FLWO 1.5m/TRES} &   \colhead{SMARTS 1.5m/CHIRON} & \colhead{SED (Stassun)} &  \colhead{SED (EXOFASTv2)}
}
\startdata
\textbf{\targetA:}
\\
~~~~$M_{*}$ (\msun)   &   ---   &   $1.1 \pm 0.1$   &   $1.06 \pm 0.06$   &   $0.998^{+0.068}_{-0.057}$
\\
~~~~$R_{*}$ (\rsun)   &   ---   &   $1.04 \pm 0.05$   &   $1.092 \pm 0.020$   &  $1.088 \pm 0.014$
\\
~~~~$L_{*}$ (\lsun)   &   ---   &   $1.06 \pm 0.11$   &   ---   &   $1.078^{+0.028}_{-0.030}$
\\
~~~~\teff\ (K)   &   $5666 \pm 50$   &    $5780 \pm 100$   &   ---   &   $5640^{+34}_{-37}$
\\
~~~~\logg\ (cgs)   &   $4.41 \pm 0.10$   &   $4.23 \pm 0.20$   &   ---   &    $4.364^{+0.032}_{-0.028}$
\\
~~~~\vsini\ ($\mathrm{km\ s^{-1}}$)   &   $3.54 \pm 0.5$   &   ---   &   ---   &   ---
\\
~~~~\feh\ (dex)   &   $0.15 \pm 0.08$   &   $0.34 \pm 0.20$   &   ---   &   $0.143^{+0.076}_{-0.078}$
\\
~~~~Age (Gyr)   &   ---   &   ---    &    ---    &  $7.3^{+3.5}_{-3.6}$
\\
\smallskip
\\
\textbf{\targetB:}
\\
~~~~$M_{*}$ (\msun) &   ---   &   $0.85 \pm 0.10$   &   $1.15 \pm 0.07$   &   $0.968^{+0.061}_{-0.068}$
\\
~~~~$R_{*}$ (\rsun)   &   ---   &   $1.14 \pm 0.03$   &   $0.964 \pm 0.052$   &  $0.918^{+0.038}_{-0.036}$ 
\\
~~~~$L_{*}$ (\lsun)   &   ---   &   $0.93 \pm 0.05$   &   ---   &   $0.730^{+0.12}_{-0.095}$
\\
~~~~\teff\ (K)   &   ---   &    $5300 \pm 100$   &   ---   &   $5570^{+150}_{-140}$
\\
~~~~\logg\ (cgs)   &   ---   &   $3.94 \pm 0.20$   &   ---   &    $4.498^{+0.025}_{-0.027}$
\\
~~~~\vsini\ ($\mathrm{km\ s^{-1}}$)   &   ---   &   ---   &   ---   &   ---
\\
~~~~\feh\ (dex)   &   ---   &   $0.20 \pm 0.10$   &   ---   &   $0.14^{+0.22}_{-0.18}$
\\
~~~~Age (Gyr)   &   ---   &   ---    &    ---    &  $3.4^{+3.8}_{-2.3}$
\\
\enddata
\end{deluxetable*}

\subsection{Results from FLWO 1.5m/TRES}
\label{sec:stellar_tres}

We derived spectral parameters from the TRES spectra of \targetA\ using the Spectral Parameter Classification (SPC) tool \citep{Buchhave2012}. SPC cross correlates the observed spectrum against a grid of synthetic spectra based on Kurucz atmospheric models \citep{Kurucz1993}. \teff, \logg, \feh, and \vsini\ are allowed to vary as free parameters. We find that $\teff = 5666 \pm 50$~K, $\logg = 4.41 \pm 0.10$, $\feh = 0.15 \pm 0.08$, and $\vsini = 3.54 \pm 0.5$~km~$\mathrm{s^{-1}}$.

\subsection{Results from SMARTS 1.5m/CHIRON}
\label{sec:stellar_chiron}

We derive the atmospheric parameters of both \targetA\ and \targetB\ following the method presented in \citet{Jones2011} and \citet{Wang2019}. We used the CHIRON template (see \S \ref{sec:RV}) to measure the equivalent width (EW) of a total of 110 Fe\,{\sc i} and 20 Fe\,{\sc ii} lines in the weak-line regime (EW $<$ 150 \AA). The EWs were measured after fitting the local continuum using the ARES2\,v2 automatic tool \citep{Sousa2015}.

We then solved the radiative transfer equation by imposing local excitation and ionization equilibrium (Boltzmann and Saha equations, respectively) and assuming a solar metal content distribution. For this, we used the \texttt{MOOG} code \citep{Sneden1973} along with the \citet{Kurucz1993} stellar atmosphere models. For models with different \teff\,, \logg\,, and \feh\,, we iterate until no dependence between the excitation potential and wavelength of the individual lines with the model abundance is found, and with the constraint that the model abundance is the same for both the Fe\,{\sc i} and Fe\,{\sc ii} lines. We note that the microturbulence velocity ($v_t$) is a free parameter in the fit. Using this method, we obtained the following atmospheric parameters for \targetA: \teff\,= 5780 $\pm$ 100 K, $\logg = 4.23 \pm 0.20$ dex, $\feh = 0.34 \pm 0.20$ and $v_t$ = 0.75 $\pm$ 0.10 km\,s$^{-1}$. For \targetB\,, we found: \teff\,= 5300 $\pm$ 100 K, $\logg = 3.94 \pm 0.20$ dex, and $\feh = 0.20 \pm 0.10$.

We adopted a value of A$_v$ = 0.10 $\pm$ 0.10 for the interstellar reddening to derive corrected visual apparent magnitudes. We also correct the \gaia\ parallax by a systematic offset of 82 $\pm$ 32 \muarcsec\ \citep{StassunTorres2018} to obtain $\varpi$ = 5.0638 $\pm$ 0.04738 and $6.66^{+0.32}_{-0.30}$ for \targetA\ and \targetB\ respectively. Using the bolometric corrections presented in \citet{Alonso1999}, we calculate a stellar luminosity of L$_\star$ = 1.06 $\pm$ 0.11 \lsun. Finally, by comparing the L$_\star$, \teff\,, and [Fe/H] with the PARSEC evolutionary tracks \citep{Bressan2012}, we derived a stellar mass and radius of 1.1 $\pm$ 0.1 \msun\ and 1.04 $\pm$ 0.05 \rsun\ respectively for \targetA\,, and 0.85 $\pm$ 0.10 \msun\ and 1.14 $\pm$ 0.03 \rsun\ for \targetB.

\subsection{Results from independent SED fitting}
\label{sec:stellar_sed}
\begin{figure}
\vspace{0cm}\hspace{0cm}
\includegraphics[width=\columnwidth]{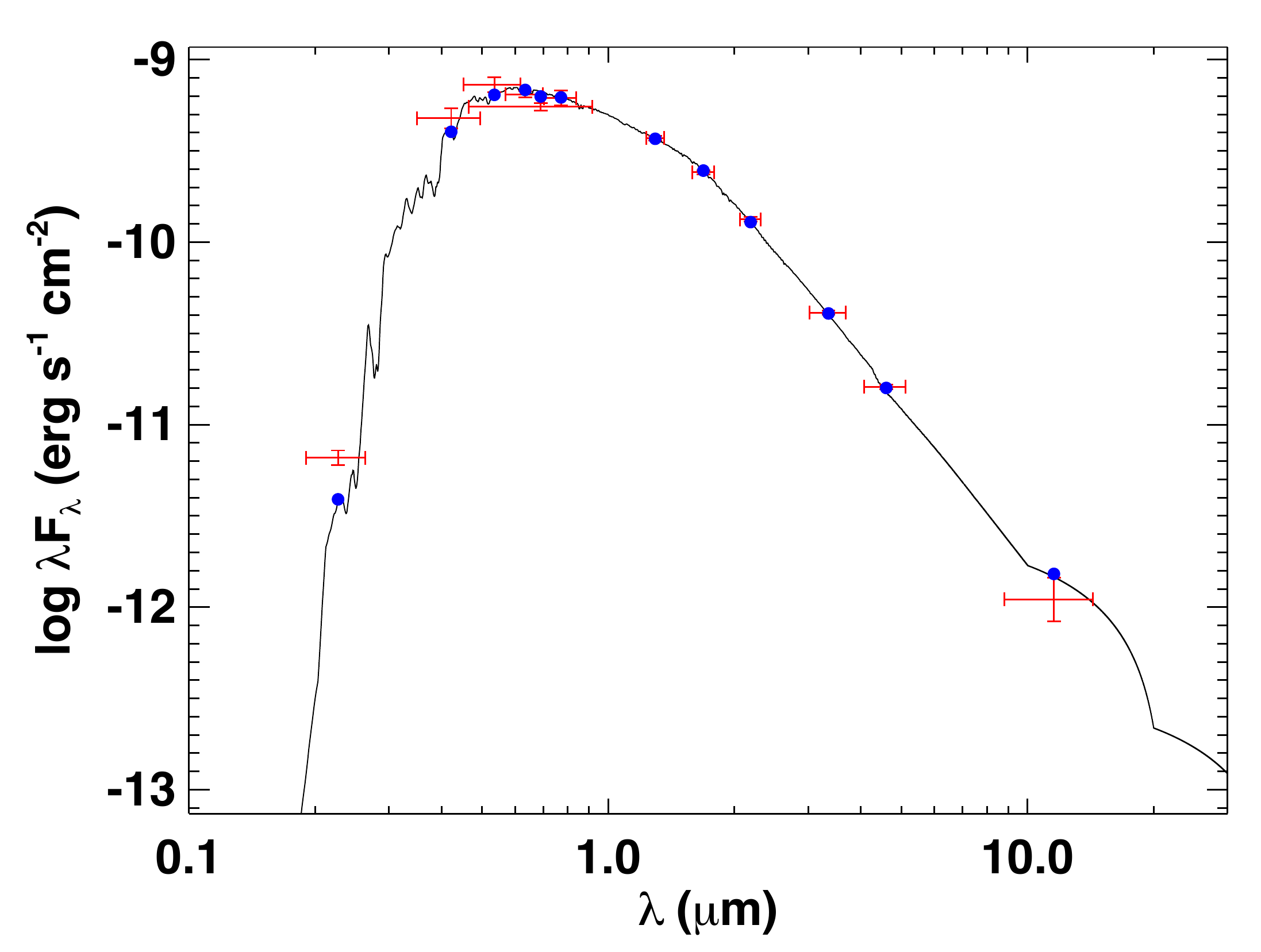}
\caption{Spectral energy distribution (SED). Red symbols represent the observed photometric measurements, where the horizontal bars represent the effective width of the passband. Blue symbols are the model fluxes from the best-fit Kurucz atmosphere model (black). 
\label{fig:sed}}
\end{figure}

Although we will compute stellar parameters based on the broadband spectral energy distribution (SED) in a global analysis using EXOFASTv2 (see \S \ref{sec:global}), we also perform a separate SED analysis as an independent check on the derived stellar parameters. Here, we use the SED together with the \gaia\ DR2 parallax in order to determine an empirical measurement of the stellar radius following the procedures described in \citet{StassunTorres2016,Stassun2017,Stassun2018a}. We pulled the $B_T V_T$ magnitudes from Tycho-2, the $BVgri$ magnitudes from APASS, the $JHK_S$ magnitudes from {\it 2MASS}, the W1--W4 magnitudes from {\it WISE}, the $G$ magnitude from \gaia, and the NUV magnitude from {\it GALEX}. Together, the available photometry spans the full stellar SED over the wavelength range 0.2--22~$\mu$m for \targetA\ and 0.4--22~$\mu$m for \targetB\ (see Figure~\ref{fig:sed}). 

We performed a fit using Kurucz stellar atmosphere models. The priors on effective temperature (\teff), surface gravity (\logg), and metallicity (\feh) were from spectroscopically determined values for \targetA\ and from the values provided in the TIC \citep{Stassun2018b} for \targetB. The remaining free parameter is the extinction ($A_V$), which we restricted to the maximum line-of-sight value from the dust maps of \citet{Schlegel1998}. The resulting fits are excellent (Figure~\ref{fig:sed}) with a reduced $\chi^2$ of 3.8 when excluding the NUV flux, which suggests mild chromospheric activity. The best-fit extinction is $A_V = 0.03_{-0.03}^{+0.11}$. Integrating the (unreddened) model SED gives the bolometric flux at Earth of $F_{\rm bol} = 9.04 \pm 0.32 \times 10^{-10}$ erg~s~cm$^{-2}$. Taking the $F_{\rm bol}$ and \teff\ together with the \gaia\ DR2 parallax, adjusted by $+0.08$~mas to account for the systematic offset reported by \citet{StassunTorres2018}, gives the stellar radius as $R = 1.092 \pm 0.020$~R$_\odot$. Finally, estimating the stellar mass from the empirical relations of \citet{Torres2010} gives $M = 1.06 \pm 0.06\ M_\odot$, which with the radius gives the mean stellar density $\rho = 1.15 \pm 0.12$ g~cm$^{-3}$.

\section{Planetary System Parameters from Global Analysis}
\label{sec:global}



We model planetary system and stellar parameters, as in \citet{Wang2019}, using EXOFASTv2\footnote{https://github.com/jdeast/EXOFASTv2} \citep{Eastman2013, Eastman2017, Eastman2019}, a fast and powerful exoplanetary fitting suite. We performed a global simultaneous analyses of both systems using light curves from \tess, LCO-CTIO, PEST, LCO-SSO, Brierfield, El-Sauce, RVs from CHIRON, and stellar spectral energy distributions. We did not include the MLAO $B$-band light curve for \targetA\ because of its lower precision compared to the LCO-SSO $B$-band light curve. We also did not include the two TRES or the two HARPS RVs, which, on their own, were not informative enough to justify introducing an additional two degrees of freedom to the fitting (namely instrumental offset and instrumental jitter); nevertheless, we note that each of these pairs of RVs were consistent with the CHIRON RVs in both systems.

During the global fitting, we applied the quadratic limb darkening law and performed a coefficients fit with a \tess-band prior based on the relation of stellar parameters  (\logg, \teff, and \feh) and coefficients \citep{Claret2018}. The corrected \gaia\ parallax for each target (\S \ref{sec:stellar_chiron}) is adopted as the Gaussian prior imposed on the \gaia\ DR2 parallaxes. An upper limit is imposed on the $V$-band extinction of 0.14 from \citet{Schlafly2011}.

To constrain each SED, we utilized the photometry from Tycho \citep{Hog2000}, 2MASS $\it JHK$ \citep{Cutri2003}, AllWISE \citep{Cutri2013}, and \gaia\ DR2 \citep{Gaia2018}; these magnitudes are given in Table \ref{tab:basic_parameters}. With the initial value of \teff\ ($5780 \pm 100$ K and $5300 \pm 100$ K for \targetA\ and \targetB\ respectively) derived from \S \ref{sec:stellar_chiron}, we utilized the available spectral energy distribution and the MIST stellar-evolutionary models \citep{Dotter2016, Choi2016} to further constrain the stellar parameters.

We began the fit with relatively standard Hot Jupiter starting conditions, but, as suggested by \citet{Eastman2019}, we iterated with relatively short MCMC runs with parallel tempering enabled to refine the starting conditions and ensure that the AMOEBA optimizer could find a good solution to all constraints simultaneously. This is not strictly required, but can dramatically improve the efficiency of EXOFASTv2. Once we found a good solution, we ran a final fit until the standard criteria (both the number of independent draws
being greater than 1000 and a Gelman-Rubin statistic of less than 1.01 for all parameters) were satisfied six consecutive times, indicating that the chains were considered to be well-mixed \citep{Eastman2013}.

Table \ref{tab:exofast} summarizes the relevant parameters reported by EXOFASTv2, with median value and 68\% confidence intervals (CI) for each posterior. \targetA\ is found to be Sun-like, with a mass of $0.998^{+0.068}_{-0.057}$ \msun, radius of $1.088 \pm 0.014$ \rsun, and \teff\ of $5640^{+34}_{-37}$~K. \targetB\ is slightly smaller with a mass of $0.968^{+0.061}_{-0.068}$ \msun, radius of $0.918^{+0.038}_{-0.036}$ \rsun, and \teff\ of $5570^{+150}_{-140}$. The two stars are each metal-rich with $\feh = 0.143^{+0.076}_{-0.078}$ and $0.14^{+0.22}_{-0.18}$ dex, respectively, which is consistent with our understanding of hot Jupiter host stars \citep{FischerValenti2005}.

The masses of both planets are determined from the CHIRON RVs and the modeled inclinations. \targetAb\ has a mass of $1.463^{+0.10}_{-0.096} \mj$, and \targetBb\ has a mass of $0.667^{+0.042}_{-0.041} \mj$. The RV curves corresponding to the median posterior values for the relevant orbital and planetary parameters are shown in Figure \ref{fig:rv_plot564} and Figure \ref{fig:rv_plot905} in black.

The transit models based on the median posterior values for each planet and photometric band are plotted in Figure \ref{fig:photometry564} and Figure \ref{fig:photometry905}. EXOFASTv2 finds a median radius and 68\% CI of \targetBb\ of $1.171^{+0.053}_{-0.051} \rj$. The radius of \targetAb\ is far more difficult to constrain; we find a median and 68\% CI of $1.02^{+0.71}_{-0.29}$ \rj. This value is very sensitive to small changes in the the impact parameter, which we determine to be $0.994^{+0.083}_{-0.049}$ with an inclination of $78.38^{+0.71}_{-0.85}$ degrees. This high impact parameter corresponds to a grazing transit scenario, and it creates a tricky interplay between the modeled $R_{p}$ and $b$.

The uncertainties in the radius of \targetAb\ are compounded in the median and 68\% CI of the bulk density estimate of $1.7^{+3.1}_{-1.4}$~g~cm$^{-3}$. The density of \targetBb\ is more precisely determined to be $0.515^{+0.063}_{-0.057}$~g~cm$^{-3}$. We find that the eccentricity is consistent with 0 with a median value and 68\% CI of $0.072^{+0.083}_{-0.050}$. Indeed, a circular orbit is to be expected for this planet based on the rapid tidal circulation timescale \citep[as computed by][with $Q_*=10^6$]{AdamsLaughlin2006} of $0.043^{+0.20}_{-0.040}$~Gyr, which is very short compared to the stellar age of $7.3^{+3.6}_{-3.5}$~Gyr. \targetBb\ also has an eccentricity that is consistent with a circular orbit: $0.024^{+0.025}_{−0.017}$. The tidal circularization timescale for this planet is $0.323^{+0.063}_{−0.054}$~Gyr.

\section{Discussion}
\label{sec:discussion}

\subsection{Consideration of False Positive Scenarios}
\label{ss:false_pos}

False positive scenarios such as background eclipsing binaries, or nearby eclipsing binaries can masquerade as giant planets in transit data and occasionally also in radial velocity data as well \citep[e.g.,][]{Torres2004}. We should be especially wary of these scenarios given that nearby stars were detected for both of these primary host stars. Here we explicitly discuss several tests for false positive scenarios. Taken together, we find that the lines of evidence collectively demonstrate the planetary nature of these bodies.

\subsubsection{RV CCF Correlations}
\label{sss:ccf_corr}

\begin{figure}
\vspace{0cm}\hspace{0cm}
\includegraphics[width=\columnwidth]{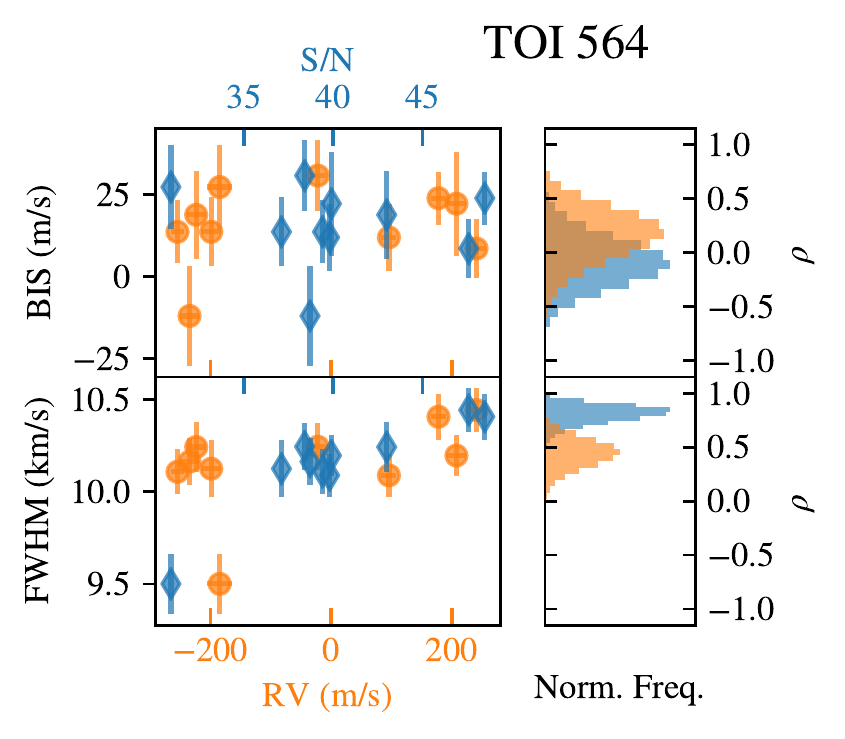}
\caption{CCF correlations for the CHIRON observations of \targetA. Left: BIS (top) and the FWHM (bottom) of the CCF are each plotted vs. both RV (orange circles) and vs. S/N (blue diamonds). Right: histograms of the Pearson correlation coefficient ($\rho$) values between either BIS (top) or FWHM (bottom) vs. RV (orange) or S/N (blue), based on a resampling of the data on the left.
\label{fig:rv_corr_564}}
\end{figure}

\begin{figure}
\vspace{0cm}\hspace{0cm}
\includegraphics[width=\columnwidth]{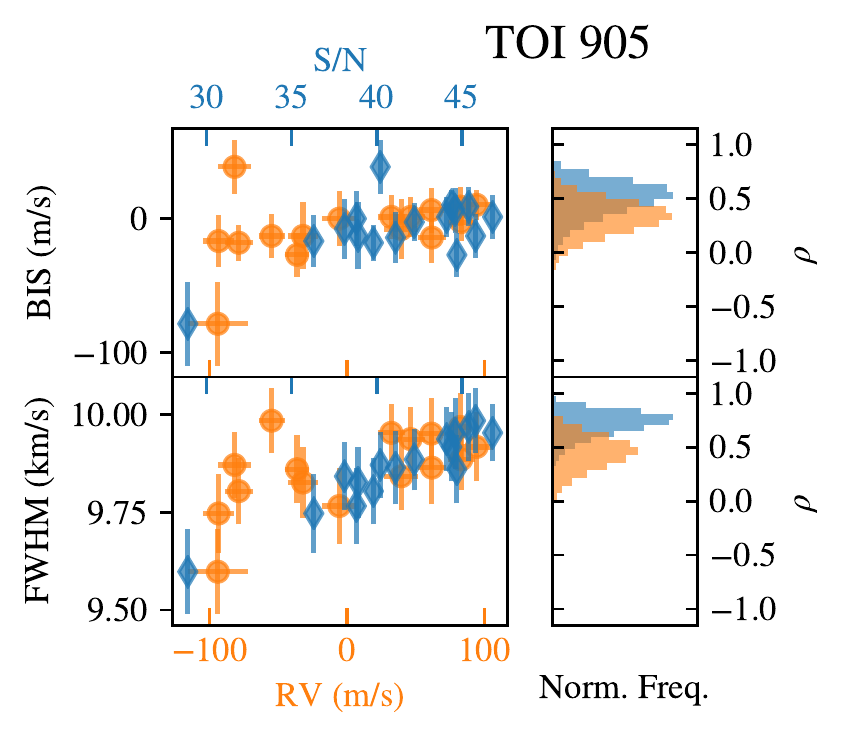}
\caption{Same as Figure \ref{fig:rv_corr_564}, but for \targetB.
\label{fig:rv_corr_905}}
\end{figure}

Strong correlations between the CCF BIS or CCF FWHM and RV can raise concerns of stellar activity masquerading as planet signals. Additionally, a marginally resolved double-lined binary can cause RV-correlated CCF variations while producing a false positive diluted planetary transit signal.

We examined the BIS and FWHM for possible linear correlations with both the RVs using the Pearson correlation coefficient, $\rho$. We calculated $\rho$ over 100,000 realizations of the data resampled from a bivarate normal distribution using the $1\sigma$ errors in both quantities. The results are seen in orange in Figure \ref{fig:rv_corr_564} and Figure \ref{fig:rv_corr_905} for \targetA\ and \targetB\ respectively. \targetA\ shows no correlations between BIS and RV, but the zero correlation case for FWHM and RV is excluded with high confidence. For \targetB, there is a marginally significant correlation between BIS and RV, and, again, a highly significant non-zero correlation for FWHM and RV.

These correlations are potentially concerning. However, we believe that the correlations are better explained as manifestations of systematic errors in our reduction pipeline that increase with low S/N. The correlations between the CCF and S/N are also in Figure \ref{fig:rv_corr_564} and Figure \ref{fig:rv_corr_905} plotted in blue. We find that in each of the cases where zero correlation was ruled out between the RVs and BIS or FWHM, the correlation was many-$\sigma$ stronger when compared to S/N instead. For instance, for \targetB\ the FWHM and RV have non-zero correlation at the $3.3\sigma$-level, whereas the FWHM and S/N have a non-zero correlation at the $9.5\sigma$-level. Therefore, we do not conclude that we have observed statistically significant astrophysically produced CCF-RV correlations, and we so fail to refute the planetary interpretation.

\subsubsection{TODCOR Analysis}
\label{sss:todcor}
Since a potentially significant bisector correlation was detected in the CCFs of \targetB, we further analyzed these spectra using TODCOR \citep{ZuckerMazeh1994}. We searched for additional RV components separated by less than $\sim$15 km/s in a procedure similar to the one applied in the analysis of the wide binary companion of HD 202772 \citep{Wang2019}. The search revealed no significant secondary velocity signal. TODCOR confirmed the RV signal was on target, and that it was not induced by a blend with another component. The upper limit on the relative flux contribution of another star in the system was estimated to be $\sim$5--10\%.

An independent reduction of the RVs obtained with TODCOR and BIS measurements obtained with UNICOR \citep{Engel2017} reproduced the observed RV semi-amplitude using the reduction described in \S \ref{ss:CHIRON}, but not the strong CCF correlations found by the CHIRON reduction; this reinforces our conclusion that the correlations discussed in \S \ref{sss:ccf_corr} are dominated by reduction issues and are not astrophysical in origin.

\subsubsection{Bounds for Inclination of M-dwarf Binary}
\label{sss:m_binary}

The apparent grazing transit of \targetA\ combined with the existence of a likely M-dwarf companion in the system raises the possibility that the system may consistent of a close M-dwarf binary pair that orbit the G star in a hierarchical triple system. In this scenario, an eclipse of the M-dwarf pair would be contaminated by the bright G star, leading to a spurious planetary transit signal, and the RVs of the system would similarly consist of high amplitude RVs from the M-dwarf pair that are diluted by the G star.

The agreement of the 1.65 day period between the transit and RV observations rule out a mutual-eclipse scenario in a hypothetical M-dwarf pair; such a binary must only experience one eclipse per orbit. A single-eclipse orbit necessitates an orbit that has non-zero eccentricity. The RVs for this system constrain the eccentricity to $0.072^{+0.083}_{-0.050}$.

We simulated this geometry, assuming the two M-dwarfs each were 0.3 \msun\ and 0.3 \rsun. The semi-major axis is therefore 0.0231 AU. We find that an inclination of $82.5^\circ < i < 83.5^\circ$ is required to produce exactly one eclipse in this scenario. Although possible, it would be unlikely \textit{a priori} to find a two-body system that falls within such a tight inclination bounds.

\subsubsection{Color Dependence}
\label{sss:color_dependence}
In the hierarchical triple system scenario for \targetA\ explored in \S \ref{sss:m_binary}, an eclipse of an M dwarf would produce transits that are deeper in redder wavelengths, in accordance with the cool stars' colors. A typical M3 star has $B-V = 1.5$, compared to the primary star's measured $B-V$ value of 0.77. This color difference between the two spectral types corresponds to a factor of $\sim$2 difference in the expected transit depths between $B$ and $V$ if the M dwarf is being eclipsed. Instead, we find that there is only a $\sim$ 0.1 ppt depth difference between the transit depth in $B$ and $V$ for this star; indeed, all transit depths only vary by 0.5 ppt between $B$ and $z'$.

The hierarchical triple system scenario would similarly cause a color dependence in the RVs, due to a varying amount of spectral contamination from the red to the blue wavelengths. We found that RVs as a function of spectral order were distributed randomly, and there is no sign of any color dependence.


\subsection{Planets in Context}
\label{ss:planet_context}

\begin{figure*}
\vspace{0cm}\hspace{0cm}
\includegraphics[width=\linewidth]{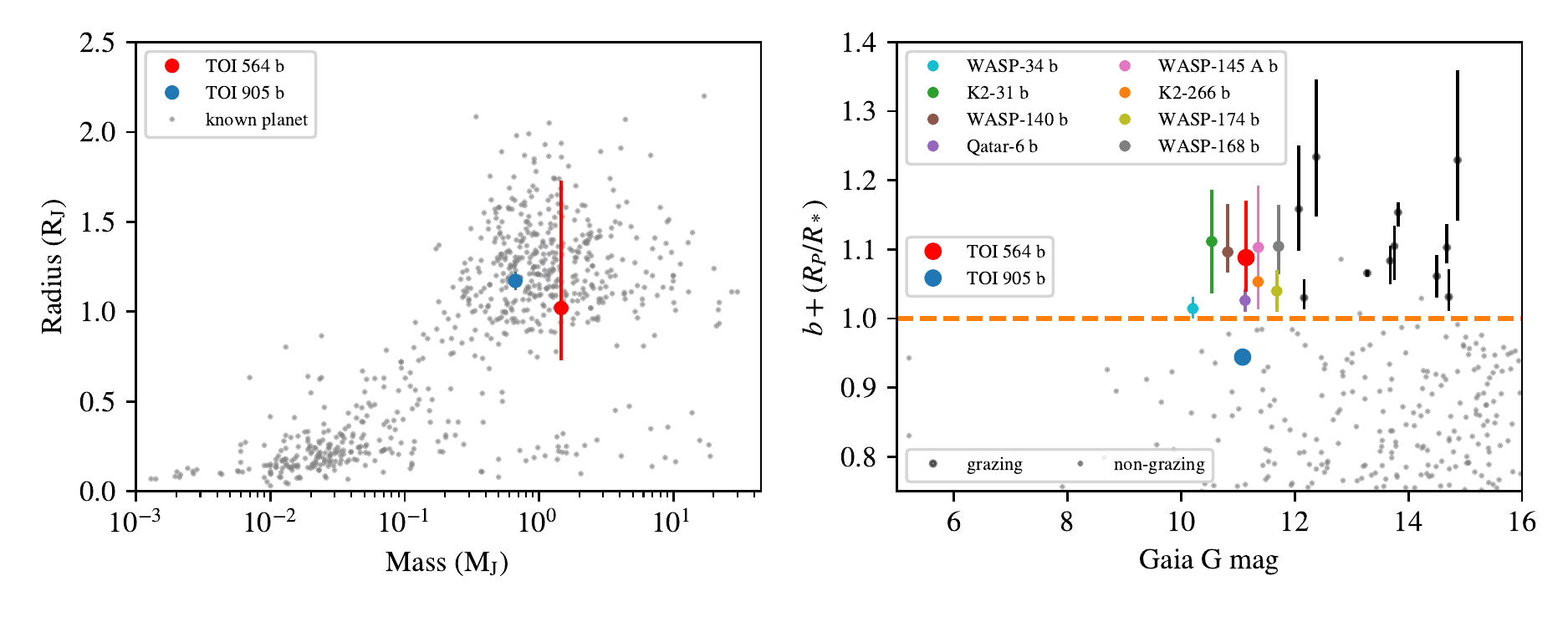}
\caption{Newly discovered planets \targetAb\ and \targetBb\ compared to known planets that have parameters published on the NASA Exoplanet Archive \citep{Akeson2013}. Left: Mass-radius relation for confirmed transiting planets, with \targetAb\ and \targetBb\ in blue and red, respectively, with 68\% confidence indicated (errors omitted for known planets). The both masses and radii are typical of gas giant planets. Right: The grazing transit condition vs. $G$ (\gaia) magnitude. The orange dashed line indicates the threshold above which transits become grazing. Planets with grazing transit probability greater than 84\% (i.e., based on published $1\sigma$ uncertainties) are labeled in various colors or in black. \targetA\ is among the brightest stars known to host a grazing transiting planet with high confidence.
\label{fig:mass_grazing}}
\end{figure*}

\targetAb\ and \targetBb\ are each classical transiting hot Jupiters, orbiting G-type host stars with (respectively) periods of 1.65 d and 3.74 d, masses of 1.46 and 0.67 \mj, and radii of 1.02 and 1.17 \rj. Figure \ref{fig:mass_grazing} (left) shows that these planets' masses and radii compared to other transiting planets. \targetBb\ sits comfortably among previously discovered gas giants, as does \targetAb, even near the extremes of its 68\% CI radius values. Given the nature of the grazing transit, however, \targetAb\ could be much larger (commensurate with a greater value of the impact parameter).

Based on \targetAb's calculated $T_{\mathrm{eq}}$ of $1714^{+20}_{-21}$~K, the fact this is a gas-giant mass planet, and comparison to other known hot Jupiters \citep[see, e.g.,][]{Wu2018}, it is likely that \targetAb\ is inflated. A typical hot Jupiter at this $T_{eq}$ would have a radius of $\sim$1.3 \rj, which is consistent within our measured radius of $1.02^{+0.71}_{-0.29}$ \rj. If \targetAb's radius fell at the low end of the 68\% CI derived from EXOFASTv2, it would be one of the least inflated giant planets given its temperature. Given the difficulties in modeling a grazing transit, we suggest that \targetAb\'s radius should be viewed cautiously as a lower-bound. \targetBb\ has a calculated $T_{\mathrm{eq}}$ of $1192^{+39}_{-36}$~K, which puts it just past the critical temperature of inflation ($1123.7 \pm 3.3$~K) found by \citet{Wu2018}. At 1.17 \rj, this planet is fully consistent with known giant planets at this $T_{\mathrm{eq}}$.

The high impact parameter, $b = 0.994^{+0.083}_{-0.049}$, makes \targetAb\ stand out in Figure \ref{fig:mass_grazing} (right), which plots the grazing transit condition vs. $Gaia$ magnitude. \targetA\ is among the brightest stars known to host a grazing transiting planet. This makes it one of the most attractive targets for long term monitoring in searches for transit depth variations and impact parameters variations that could reveal the presence of non-transiting planets or exomoons \citep{Kipping2009, Kipping2010}. The grazing transits also offer an opportunity to search for exotrojan asteroids \citep[e.g.,][]{LilloBox2018} by exploiting the sensitivity of the orbit of \targetAb\ to co-orbital perturbations.

\citet{MiraldaEscude2002} examines this possibility using the 51 Peg system \citep{MayorQueloz1995} as an example. A close-in hot Jupiter will experience both a precession in its periastron as well as a precession in its line of nodes when an additional planet is present in the system. \citet{MiraldaEscude2002} finds, for example, that in the case of an Earth-mass planet located at $a = 0.2$~AU with an inclination of $45^\circ$, 51 Peg b would experience transit duration changes of 1~s~$\mathrm{yr^{-1}}$, which would be detectable over many years of observation. However, the grazing nature of \targetAb's transit means that any line-of-nodes precession will manifest itself as a change to the already-high impact parameter, upon which the transit duration and transit depth are both extremely sensitive. These changes can be used to dynamically constrain the presence of smaller and more distant planetary companions, which are of particular interest because they can help address hot Jupiter formation and evolution scenarios.


We find that \targetAb\ is unlikely to be eclipsed by its host star, with the 68\% CI eclipse duration being $0.000^{+0.021}_{-0.00}$ d. However, \targetBb\ is expected to be eclipsed by its host, with an eclipse duration of $0.0845^{+0.0011}_{-0.0015}$ d. There is therefore potential for further study of this planet's thermal emission and atmospheric characterization.


We simulated the Rossiter-McLaughlin effect using ExOSAM \citep[see,][]{Addison2018} for both \targetAb\ and \targetBb. Following the results of the TRES observations, we set set $\vsini = \mathrm{3.5\ km\ s^{-1}}$ for \targetA, and we consider two cases: an aligned orbit with $\lambda = 0^{\circ}$, and with a polar orbit with $\lambda = 90^{\circ}$. For an aligned orbit, the predicted semi-amplitude of the velocity anomaly is $\mathrm{\sim 8\ m\ s^{-1}}$. In the case where $\lambda \sim 90^{\circ}$ and the star has a non-trivial vsini, either the red-shifted or the blue-shifted limb of the star will be occulted, which will result in a measurable, fully asymmetric Rossiter-McLaughlin signal of $\mathrm{\sim 7\ m\ s^{-1}}$. With a transit duration of just over one hour, it would be readily possible for a large-aperture telescope with a spectrograph attaining better than $\mathrm{4\ m\ s^{-1}}$ precision with a cadence of $\sim10$\,min, (e.g., Keck/HIRES, \citealt{Vogt1994}; Magellan/PFS, \citealt{Crane2006}), to measure $\lambda$ in this system. We do not have measurements of \vsini\ for \targetB, but the deeper transit would probably create a larger RM signal; e.g., arbitrarily assuming the same \vsini\ of $\mathrm{3.5\ km\ s^{-1}}$, we find that \targetBb\ would have an RM semi-amplitude of $\mathrm{\sim 23\ m\ s^{-1}}$ for an aligned orbit.

The confirmation of this grazing transiting planet may also serve as a valuable data point in informing the mystery behind the paucity in detections of exoplanets with grazing transits, even after accounting for the detection biases resulting from their shallower and shorter transits. Polar star spots have been observed on both main sequence stars \citep{Jeffers2002} and active, rapid rotators \citep{SchuesslerSolanki1992}. These spots reduce the background flux of the region occulted by planets exhibiting a grazing transit, which necessarily transit at high latitude in the default case of $\lambda = 0^{\circ}$. \citet{Oshagh2015} posits that this effect could be responsible for the dearth of grazing transiting planet detections by Kepler. If \targetAb\ is indeed in an aligned orbit, then its transits must cross the stellar pole (because of the grazing transits), which would grant us the opportunity to study this phenomenon.

\section{Summary and Conclusion}

We report the discovery and confirmation of two new hot Jupiters identified by \tess: \targetAb\ and \targetBb. \targetAb\ is noteworthy in that it displays a grazing transit across its Sun-like host star in over its 1.65 d orbit. Both targets are main sequence G stars that are relatively bright ($V\sim11$), making them good targets for follow-up characterization.

Both planets were validated based on the \tess\ light curves, ground-based photometry in three different filters, and robust radial velocity detections by CHIRON. Both stars were observed with speckle interferometry (HRCam/SOAR) and \targetA\ was also observed with PHARO/Palomar AO. \targetA\ is a probable binary system, with an M-dwarf companion at a projected distance of $\sim$100 AU.

We conducted multiple independent measurements of the host stars' stellar parameters using the high resolution CHIRON and TRES spectra as well as an SED analysis and found a general agreement between the derived parameters. Using the EXOFASTv2 planet fitting suite, we ran a global analysis by simultaneously fitting the transit and RV data with a Markov chain Monte Carlo. \targetAb's impact parameter was found to be near unity, diminishing our ability to constrain its radius, but its mass, as well as the radius and mass of \targetBb, was measured with high precision.

We explored and rejected a variety of false positive scenarios for both systems. We conducted simulations of the Rossiter-McLaughlin effect for \targetAb, assuming either an equatorial orbit or a polar orbit and found that in both cases the RV anomaly ($\sim 8\ \mathrm{m\ s^{-1}}$) should be detectable by large aperture telescopes with spectrographs capable of attaining $\sim \mathrm{4\ m\ s^{-1}}$ precision with $\sim 10$ minute cadence. We noted that the unique sensitivity of grazing transits to small orbital perturbations like inclination changes (the effects of which are magnified when the transit depth changes) may be exploited to search the system for additional non-transiting bodies. 

\acknowledgments

A.B.D. is supported by the National Science Foundation Graduate Research Fellowship Program under Grant Number DGE-1122492. S.W. thanks the Heising-Simons Foundation for their generous support as a 51 Pegasi b fellow. M.N.G. acknowledges support from MIT’s Kavli Institute as a Torres postdoctoral fellow. C.Z. is supported by a Dunlap Fellowship at the Dunlap Institute for Astronomy \& Astrophysics, funded through an endowment established by the Dunlap family and the University of Toronto. Work by J.N.W. was partly supported by the Heising-Simons Foundation. 

Funding for the TESS mission is provided by National Aeronautics and Space Administration's (NASA) Science Mission directorate. We acknowledge the use of public \tess\ Alert data from pipelines at the \tess\ Science Office and at the \tess\ Science Processing Operations Center.
Resources supporting this work were provided by the NASA High-End Computing (HEC) Program through the NASA Advanced Supercomputing (NAS) Division at Ames Research Center for the production of the SPOC data products.
This research has made use of the Exoplanet Follow-up Observation Program website and the NASA Exoplanet Archive, which is operated by the California Institute of Technology, under contract with the National Aeronautics and Space Administration under the Exoplanet Exploration Program.
This paper includes data collected by the TESS mission, which are publicly available from the Mikulski Archive for Space Telescopes (MAST).

Part of this research was carried out at the Jet Propulsion Laboratory, California Institute of Technology, under a contract with NASA.

This work makes use of observations from SMARTS and the LCOGT network.

We made use of the Python programming language \citep{Rossum1995}
and the open-source Python packages
\textsc{numpy} \citep{VanderWalt2011}, 
\textsc{emcee} \citep{Foreman-Mackey2013}, and
\textsc{celerite} \citep{Foreman-Mackey2017}.


\facilities{TESS, CTIO:1.5m (CHIRON), FLWO:1.5m (TRES), ESO:3.6m (HARPS), LCOGT, SOAR (HRCam), Hale (PHARO)}



\begin{longrotatetable}
\begin{deluxetable*}{lccccc}
\tablecaption{Median Values and 68\% Confidence Interval for the \targetA\ and \targetB\ Planetary Systems\label{tab:exofast}}
\tablehead{\colhead{~~~Parameter} & \colhead{Units} & \multicolumn{4}{c}{Values}}
\startdata
\smallskip\\\multicolumn{2}{l}{Stellar Parameters:}&\targetA&\targetB&\smallskip\\
~~~~$M_*$\dotfill &Mass (\msun)\dotfill &$0.998^{+0.068}_{-0.057}$&$0.968^{+0.061}_{-0.068}$\\
~~~~$R_*$\dotfill &Radius (\rsun)\dotfill &$1.088\pm0.014$&$0.918^{+0.038}_{-0.036}$\\
~~~~$L_*$\dotfill &Luminosity (\lsun)\dotfill &$1.078^{+0.028}_{-0.030}$&$0.730^{+0.12}_{-0.095}$\\
~~~~$\rho_*$\dotfill &Density (cgs)\dotfill &$1.095^{+0.090}_{-0.075}$& $1.76\pm0.16$\\
~~~~$\log{g}$\dotfill &Surface gravity (cgs)\dotfill &$4.364^{+0.032}_{-0.028}$&$4.498^{+0.025}_{-0.027}$\\
~~~~$T_{\rm eff}$\dotfill &Effective Temperature (K)\dotfill &$5640^{+34}_{-37}$&$5570^{+150}_{-140}$\\
~~~~$[{\rm Fe/H}]$\dotfill &Metallicity (dex)\dotfill &$0.143^{+0.076}_{-0.078}$&$0.14^{+0.22}_{-0.18}$\\
~~~~$[{\rm Fe/H}]_{0}$\dotfill &Initial Metallicity \dotfill &$0.165^{+0.069}_{-0.072}$&$0.12^{+0.19}_{-0.16}$\\
~~~~$Age$\dotfill &Age (Gyr)\dotfill &$7.3^{+3.6}_{-3.5}$&$3.4^{+3.8}_{-2.3}$\\
~~~~$A_V$\dotfill &V-band extinction (mag)\dotfill &$0.108^{+0.021}_{-0.033}$&$0.23\pm0.12$\\
~~~~$\sigma_{SED}$\dotfill &SED photometry error scaling \dotfill &$1.27^{+0.42}_{-0.27}$&$1.54^{+0.49}_{-0.32}$\\
~~~~$\varpi$\dotfill &Parallax (mas)\dotfill &$5.067\pm0.046$&$6.66^{+0.32}_{-0.30}$\\
~~~~$d$\dotfill &Distance (pc)\dotfill &$197.4\pm1.8$&$150.2^{+7.2}_{-6.9}$\\
\smallskip\\\multicolumn{2}{l}{Planetary Parameters:}&\targetAb&\targetBb\smallskip\\
~~~~$P$\dotfill &Period (days)\dotfill &$1.651144\pm0.000018$&$3.739494\pm0.000038$\\
~~~~$R_P$\dotfill &Radius (\rj)\dotfill &$1.02^{+0.71}_{-0.29}$&$1.171^{+0.053}_{-0.051}$\\
~~~~$M_P$\dotfill &Mass (\mj)\dotfill &$1.463^{+0.10}_{-0.096}$&$0.667^{+0.042}_{-0.041}$\\
~~~~$\rho_P$\dotfill &Density (cgs)\dotfill &$1.7^{+3.1}_{-1.4}$&$0.515^{+0.063}_{-0.057}$\\
~~~~$T_C$\dotfill &Time of conjunction (\bjdtdb)\dotfill &$2458518.20381^{+0.00057}_{-0.00058}$&$2458628.35101\pm0.00026$\\
~~~~$T_0$\dotfill &Optimal conjunction Time (\bjdtdb)\dotfill &$2458549.57554^{+0.00045}_{-0.00046}$&$2458643.30898\pm0.00020$\\
~~~~$a$\dotfill &Semi-major axis (AU)\dotfill &$0.02734^{+0.00061}_{-0.00053}$&$0.04666^{+0.00096}_{-0.0011}$\\
~~~~$i$\dotfill &Inclination (Degrees)\dotfill &$78.38^{+0.71}_{-0.85}$&$85.68^{+0.22}_{-0.26}$\\
~~~~$e$\dotfill &Eccentricity \dotfill &$0.072^{+0.083}_{-0.050}$&$0.024^{+0.025}_{-0.017}$\\
~~~~$\omega_*$\dotfill &Argument of Periastron (Degrees)\dotfill &$94^{+32}_{-35}$&$39^{+61}_{-82}$\\
~~~~$K$\dotfill &RV semi-amplitude (m/s)\dotfill &$247\pm13$&$89.1^{+3.8}_{-3.6}$\\
~~~~$T_{eq}$\dotfill &Equilibrium temperature (K)\dotfill &$1714^{+20}_{-21}$&$1192^{+39}_{-36}$\\
~~~~$\tau_{\rm circ}$\dotfill &Tidal circularization timescale (Gyr)\dotfill &$0.043^{+0.20}_{-0.040}$&$0.323^{+0.063}_{-0.054}$\\
~~~~$\delta$\dotfill &Transit depth (fraction)\dotfill &$0.0092^{+0.017}_{-0.0045}$&$0.01718^{+0.00032}_{-0.00030}$\\
~~~~$Depth$\dotfill &Flux decrement at mid transit \dotfill &$0.00484^{+0.00039}_{-0.00047}$&$0.01718^{+0.00032}_{-0.00030}$\\
~~~~$b$\dotfill &Transit Impact parameter \dotfill &$0.994^{+0.083}_{-0.049}$&$0.816^{+0.010}_{-0.012}$\\
~~~~$\tau$\dotfill &Ingress/egress transit duration (days)\dotfill &$0.02139^{+0.00062}_{-0.00077}$&$0.0278^{+0.012}_{-0.0038}$\\
~~~~$T_{14}$\dotfill &Total transit duration (days)\dotfill &$0.0428^{+0.0012}_{-0.0013}$&$0.0845^{+0.0011}_{-0.0015}$\\
~~~~$b_S$\dotfill &Eclipse impact parameter \dotfill &$1.152^{+0.14}_{-0.089}$&$0.827^{+0.047}_{-0.031}$\\
~~~~$\tau_S$\dotfill &Ingress/egress eclipse duration (days)\dotfill &$0.000^{+0.021}_{-0.00}$&$0.0278^{+0.012}_{-0.0038}$\\
~~~~$T_{S,14}$\dotfill &Total eclipse duration (days)\dotfill &$0.000^{+0.041}_{-0.00}$&$0.0845^{+0.0011}_{-0.0015}$\\
~~~~$logg_P$\dotfill &Surface gravity \dotfill &$3.55^{+0.30}_{-0.46}$&$3.081\pm0.035$\\
~~~~$\fave$\dotfill &Incident Flux (\fluxcgs)\dotfill &$1.937^{+0.099}_{-0.11}$&$0.458^{+0.063}_{-0.052}$\\
~~~~$T_P$\dotfill &Time of Periastron (\bjdtdb)\dotfill &$2458516.57^{+0.13}_{-0.14}$&$2458627.86^{+0.60}_{-0.85}$\\
~~~~$T_S$\dotfill &Time of eclipse (\bjdtdb)\dotfill &$2458517.374^{+0.028}_{-0.036}$&$2458630.245^{+0.049}_{-0.031}$\\
~~~~$T_A$\dotfill &Time of Ascending Node (\bjdtdb)\dotfill &$2458517.822^{+0.044}_{-0.030}$&$2458627.436^{+0.046}_{-0.026}$\\
~~~~$T_D$\dotfill &Time of Descending Node (\bjdtdb)\dotfill &$2458518.580^{+0.036}_{-0.050}$&$2458629.288^{+0.032}_{-0.029}$\\
~~~~$e\cos{\omega_*}$\dotfill & \dotfill &$-0.004^{+0.026}_{-0.033}$&$0.010^{+0.020}_{-0.013}$\\
~~~~$e\sin{\omega_*}$\dotfill & \dotfill &$0.063^{+0.086}_{-0.058}$&$0.006^{+0.028}_{-0.016}$\\
~~~~$d/R_*$\dotfill &Separation at mid transit \dotfill &$5.05^{+0.33}_{-0.46}$&$10.85^{+0.47}_{-0.59}$\\
\smallskip\\\multicolumn{2}{l}{Telescope Parameters:}&CHIRON (\targetA)&CHIRON (\targetB)\smallskip\\
~~~~$\gamma_{\rm rel}$\dotfill &Relative RV Offset (m/s)\dotfill &$22\pm11$&$-36.2^{+2.7}_{-2.9}$\\
~~~~$\sigma_J$\dotfill &RV Jitter (m/s)\dotfill &$25.9^{+17}_{-9.7}$&$2.0^{+6.8}_{-2.0}$\\
~~~~$\sigma_J^2$\dotfill &RV Jitter Variance \dotfill &$670^{+1100}_{-410}$&$3^{+73}_{-38}$\\
\smallskip\\\multicolumn{2}{l}{Wavelength Parameters for \targetA:}&TESS&LCO-SSO (B)&PEST (V)&LCO-CTIO (z')\smallskip\\
~~~~$u_{1}$\dotfill &Linear limb-darkening coeff \dotfill &$0.370^{+0.045}_{-0.046}$&$0.664\pm0.047$&$0.488\pm0.050$&$0.267\pm0.047$\\
~~~~$u_{2}$\dotfill &Quadratic limb-darkening coeff \dotfill &$0.301^{+0.046}_{-0.047}$&$0.079\pm0.048$&$0.220\pm0.050$&$0.265\pm0.048$\\
\smallskip\\\multicolumn{2}{l}{Wavelength Parameters for \targetB:}&TESS&LCO-SSO (G)& El Sauce ($R_\mathrm{c}$) & Brierfield (I)\smallskip\\
~~~~$u_{1}$\dotfill &Linear limb-darkening coeff \dotfill &$0.345^{+0.054}_{-0.055}$&$0.637^{+0.066}_{-0.067}$&$0.424^{+0.059}_{-0.060}$&$0.315^{+0.055}_{-0.056}$\\
~~~~$u_{2}$\dotfill &Quadratic limb-darkening coeff \dotfill &$0.263^{+0.050}_{-0.049}$&$0.153^{+0.062}_{-0.061}$&$0.262^{+0.053}_{-0.052}$&$0.253\pm0.051$\\
\smallskip\\\multicolumn{2}{l}{Transit Parameters for \targetA:}&TESS& LCO-SSO (B)&PEST (V)&LCO-CTIO (z')
\smallskip\\
~~~~$\sigma^{2}$\dotfill &Added Variance \dotfill &$0.000000010\pm0.000000024$&$0.00000091^{+0.00000025}_{-0.00000021}$&$0.0000010^{+0.0000015}_{-0.0000012}$&$0.00000052^{+0.00000016}_{-0.00000014}$\\
~~~~$F_0$\dotfill &Baseline flux \dotfill &$1.000011^{+0.000012}_{-0.000013}$&$1.00070\pm0.00022$&$0.99954\pm0.00062$&$0.99994\pm0.00019$\\
~~~~$M_{0}$\dotfill &Multiplicative detrending coeff \dotfill &--&$0.00212\pm0.00049$&$-0.0008\pm0.0013$&$-0.00027^{+0.00039}_{-0.00040}$\\
\smallskip\\\multicolumn{2}{l}{Transit Parameters for \targetB:}&TESS&LCO-SSO (G)& El Sauce ($R_\mathrm{c}$) & Brierfield (I)
\smallskip\\
~~~~$\sigma^{2}$\dotfill &Added Variance \dotfill &$0.000000004\pm0.000000032$&$0.0000064^{+0.0000015}_{-0.0000013}$&$0.0000083^{+0.0000012}_{-0.0000010}$&$0.0000207^{+0.0000034}_{-0.0000029}$\\
~~~~$F_0$\dotfill &Baseline flux \dotfill &$1.000002\pm0.000013$&$0.99993\pm0.00031$&$1.00112^{+0.00028}_{-0.00027}$&$1.00022\pm0.00044$\\
\enddata
\label{tab:TOI564.short.maxrp25.}
\end{deluxetable*}
\end{longrotatetable}

\clearpage

\clearpage




\clearpage
\begin{thebibliography}{10}

\bibitem[Adams \& Laughlin(2006)]{AdamsLaughlin2006} Adams, F.~C., \& Laughlin, G.\ 2006, \apj, 649, 1004

\bibitem[Addison et al.(2018)]{Addison2018} Addison, B.~C., Wang, S., Johnson, M.~C., et al.\ 2018, \aj, 156, 197

\bibitem[Akeson et al.(2013)]{Akeson2013} Akeson, R.~L., Chen, X., Ciardi, D., et al.\ 2013, \pasp, 125, 989

\bibitem[Albrecht et al.(2013)]{Albrecht2013} Albrecht, S., Winn, J.~N., Marcy, G.~W., et al.\ 2013, \apj, 771, 11

\bibitem[Almenara et al.(2013)]{Almenara2013} Almenara, J.~M., Bouchy, F., Gaulme, P., et al.\ 2013, \aap, 555, A118

\bibitem[Almenara et al.(2015)]{Almenara2015} Almenara, J.~M., Damiani, C., Bouchy, F., et al.\ 2015, \aap, 575, A71

\bibitem[Alsubai et al.(2018)]{Alsubai2018} Alsubai, K., Tsvetanov, Z.~I., Latham, D.~W., et al.\ 2018, \aj, 155, 52

\bibitem[Alonso et al.(1999)]{Alonso1999} Alonso, A., Arribas, S., \& Martinez-Roger, C. 1999, A\&A, 140, 

\bibitem[Alonso et al.(2008)]{Alonso2008} Alonso, R., Barbieri, M., Rabus, M., et al.\ 2008, \aap, 487, L5

\bibitem[Anderson et al.(2011)]{Anderson2011} Anderson, D.~R., Barros, S.~C.~C., Boisse, I., et al.\ 2011, \pasp, 123, 555

\bibitem[Anderson et al.(2012)]{Anderson2012} Anderson, D.~R., Collier Cameron, A., Gillon, M., et al.\ 2012, \mnras, 422, 1988

\bibitem[Bakos et al.(2004)]{Bakos2004} Bakos, G., Noyes, R.~W., Kov{\'a}cs, G., et al.\ 2004, \pasp, 116, 266

\bibitem[Batygin et al.(2016)]{Batygin2016} Batygin, K., Bodenheimer, P.~H., \& Laughlin, G.~P.\ 2016, \apj, 829, 114

\bibitem[Bayliss et al.(2018)]{Bayliss2018} Bayliss, D., Gillen, E., Eigm{\"u}ller, P., et al.\ 2018, \mnras, 475, 4467

\bibitem[Bean \& Seifahrt(2008)]{BeanSeifahrt2008} Bean, J.~L., \& Seifahrt, A.\ 2008, \aap, 487, L25

\bibitem[Becker et al.(2015)]{Becker2015} Becker, J.~C., Vanderburg, A., Adams, F.~C., et al.\ 2015, \apjl, 812, L18

\bibitem[B{\'e}ky et al.(2011)]{Beky2011} B{\'e}ky, B., Bakos, G. {\'A}., Hartman, J., et al.\ 2011, \apj, 734, 109

\bibitem[Bodenheimer et al.(2000)]{Bodenheimer2000} Bodenheimer, P., Hubickyj, O., \& Lissauer, J.~J.\ 2000, \icarus, 143, 2

\bibitem[Borucki et al.(2010)]{Borucki2010} Borucki, W.~J., Koch, D., Basri, G., et al.\ 2010, Science, 327, 977

\bibitem[Bressan et al.(2012)]{Bressan2012} Bressan, A., Marigo, P., Girardi, L. et al. 2012, MNRAS, 427, 127

\bibitem[Gaia Collaboration et al.(2018)]{Gaia2018} Gaia Collaboration, Brown, A.~G.~A., Vallenari, A., et al.\ 2018, \aap, 616, A1

\bibitem[Brown et al.(2013)]{Brown2013} Brown, T.~M., Baliber, N., Bianco, F.~B., et al.\ 2013, \pasp, 125, 1031

\bibitem[Bruno et al.(2018)]{Bruno2018} Bruno, G., Lewis, N.~K., Stevenson, K.~B., et al.\ 2018, \aj, 155, 55

\bibitem[Buchhave et al.(2010)]{Buchhave2010} Buchhave, L.~A., Bakos, G. {\'A}., Hartman, J.~D., et al.\ 2010, \apj, 720, 1118

\bibitem[Buchhave et al.(2012)]{Buchhave2012} Buchhave, L.~A., Latham, D.~W., Johansen, A., et al.\ 2012, \nat, 486, 375

\bibitem[Butler et al.(2004)]{Butler2004} Butler, R.~P., Vogt, S.~S., Marcy, G.~W., et al.\ 2004, \apj, 617, 580


\bibitem[Ca{\~n}as et al.(2019a)]{Canas2019a} Ca{\~n}as, C.~I., Wang, S., Mahadevan, S., et al.\ 2019a, \apjl, 870, L17

\bibitem[Ca{\~n}as et al.(2019b)]{Canas2019b} Ca{\~n}as, C.~I., Stefansson, G., Monson, A.~J., et al.\ 2019b, \apjl, 877, L29

\bibitem[Choi et al.(2016)]{Choi2016} Choi, J., Dotter, A., Conroy, C., et al.\ 2016, \apj, 823, 102

\bibitem[Claret et al.(2018)]{Claret2018} Claret, A. 2018, A\&A, 618, A20

\bibitem[Collins et al.(2017)]{Collins2017} Collins, K.~A., Kielkopf, J.~F., Stassun, K.~G., et al.\ 2017, \aj, 153, 77

\bibitem[Crane et al.(2006)]{Crane2006} Crane, J.~D., Shectman, S.~A., \& Butler, R.~P.\ 2006, \procspie, 626931

\bibitem[Crida \& Batygin(2014)]{CridaBatygin2014} Crida, A., \& Batygin, K.\ 2014, \aap, 567, A42

\bibitem[Cutri et al.(2003)]{Cutri2003} Cutri, R. M., Skrutskie, M. F., van Dyk, S., et al. \ 2003, yCat, 2246, 0C

\bibitem[Cutri et al.(2013)]{Cutri2013} Cutri, R. M., et al. \ 2013, yCat, 1322, 0Z

\bibitem[Dawson et al.(2019)]{Dawson2019} Dawson, R.~I., Huang, C.~X., Lissauer, J.~J., et al.\ 2019, \aj, 158, 65

\bibitem[Demory et al.(2009)]{Demory2009} Demory, B.-O., Gillon, M., Waelkens, C., et al.\ 2009, Transiting Planets, 424

\bibitem[Dong et al.(2014)]{Dong2014} Dong, S., Katz, B., \& Socrates, A.\ 2014, \apjl, 781, L5

\bibitem[Dotter(2016)]{Dotter2016} Dotter, A.\ 2016, \apjs, 222, 8

\bibitem[Eastman et al.(2013)]{Eastman2013} Eastman, J., Gaudi, B.~S., \& Agol, E.\ 2013, \pasp, 125, 83 

\bibitem[Eastman(2017)]{Eastman2017} Eastman, J.\ 2017, Astrophysics Source Code Library, ascl:1710.003 

\bibitem[Eastman et al.(2019)]{Eastman2019} Eastman, J.~D., Rodriguez, J.~E., Agol, E., et al.\ 2019, arXiv e-prints, arXiv:1907.09480

\bibitem[Engel et al.(2017)]{Engel2017} Engel, M., Shahaf, S., \& Mazeh, T.\ 2017, \pasp, 129, 065002

\bibitem[Fischer \& Valenti(2005)]{FischerValenti2005} Fischer, D.~A., \& Valenti, J.\ 2005, \apj, 622, 1102

\bibitem[Ford \& Rasio(2008)]{FordRasio2008} Ford, E.~B., \& Rasio, F.~A.\ 2008, \apj, 686, 621

\bibitem[Foreman-Mackey et al.(2017)]{Foreman-Mackey2017} Foreman-Mackey, D., Agola, E., Ambikasaran, S., Angus, R.\ 2017, \aj, 154, 220

\bibitem[Foreman-Mackey et al.(2013)]{Foreman-Mackey2013} Foreman-Mackey, D., Hogg, D.~W., Lang, D., Goodman, J.\ 2013, \pasp, 125, 306

\bibitem[Fressin et al.(2013)]{Fressin2013} Fressin, F., Torres, G., Charbonneau, D., et al.\ 2013, \apj, 766, 81

\bibitem[F{\H{u}}r{\'e}sz et al.(2008)]{Furesz2008} F{\H{u}}r{\'e}sz, G., Szentgyorgyi, A.~H., \& Meibom, S.\ 2008, Precision Spectroscopy in Astrophysics, 287

\bibitem[Grziwa et al.(2016)]{Grziwa2016} Grziwa, S., Gandolfi, D., Csizmadia, S., et al.\ 2016, \aj, 152, 132

\bibitem[G{\"u}nther et al.(2017)]{Guenther2017} {G{\"u}nther}, M.~N., {Queloz}, D., {Gillen}, E., et al.\ 2017, \mnras, 472, 295 

\bibitem[G{\"u}nther et al.(2018)]{Guenther2018} {G{\"u}nther}, M.~N., {Queloz}, D., {Gillen}, E., et al.\ 2018, \mnras, 478, 4720
  
  

\bibitem[Kurucz(1993)]{Kurucz1993} Kurucz, R.\ 1993, ATLAS9 Stellar Atmosphere Programs and 2 km/s grid.~Kurucz CD-ROM No.~13.~ Cambridge, Mass.: Smithsonian Astrophysical Observatory, 1993.

\bibitem[Hellier et al.(2012)]{Hellier2012} Hellier, C., Anderson, D.~R., Collier Cameron, A., et al.\ 2012, \mnras, 426, 739

\bibitem[Hellier et al.(2017)]{Hellier2017} Hellier, C., Anderson, D.~R., Cameron, A.~C., et al.\ 2017, \mnras, 465, 3693

\bibitem[H{\o}g et al.(2000)]{Hog2000} H{\o}g, E., Fabricius, C., Makarov, V. V., et al. \ 2000, A\&A, 335L, 27H

\bibitem[Holt(1893)]{Holt1893} Holt, J.~R.\ 1893, Astronomy and Astro-Physics (formerly The Sidereal Messenger), 12, 646

\bibitem[Howard et al.(2012)]{Howard2012} Howard, A.~W., Marcy, G.~W., Bryson, S.~T., et al.\ 2012, \apjs, 201, 15

\bibitem[Howell et al.(2014)]{Howell2014} Howell, S.~B., Sobeck, C., Haas, M., et al.\ 2014, \pasp, 126, 398

\bibitem[Huang et al.(2016)]{Huang2016} Huang, C., Wu, Y., \& Triaud, A.~H.~M.~J.\ 2016, \apj, 825, 98


\bibitem[Jeffers et al.(2002)]{Jeffers2002} Jeffers, S.~V., Barnes, J.~R., \& Collier Cameron, A.\ 2002, \mnras, 331, 666

\bibitem[Jenkins et al.(2016)]{Jenkins2016} Jenkins, J.~M., Twicken, J.~D., McCauliff, S., et al.\ 2016, \procspie, 99133E

\bibitem[Jensen(2013)]{Jensen2013} Jensen, E.\ 2013, Tapir: A web interface for transit/eclipse observability, ascl:1306.007

\bibitem[Jones et al.(2019)]{Jones2019} Jones, M.~I., Brahm, R., Espinoza, N., et al.\ 2019, \aap, 625, A16

\bibitem[Jones et al.(2017)]{Jones2017} Jones, M.~I., Brahm, R., Wittenmyer, R.~A., et al.\ 2017, \aap, 602, A58

\bibitem[Jones et al.(2011)]{Jones2011} Jones, M.~I., Jenkins, J.~S., Rojo, P., et al.\ 2011, \aap, 536, A71

\bibitem[Kipping(2009)]{Kipping2009} Kipping, D.~M.\ 2009, \mnras, 396, 1797

\bibitem[Kipping(2010)]{Kipping2010} Kipping, D.~M.\ 2010, \mnras, 407, 301

\bibitem[Knutson et al.(2014)]{Knutson2014} Knutson, H.~A., Fulton, B.~J., Montet, B.~T., et al.\ 2014, \apj, 785, 126

\bibitem[Kossakowski et al.(2019)]{Kossakowski2019} Kossakowski, D., Espinoza, N., Brahm, R., et al.\ 2019, arXiv e-prints, arXiv:1906.09866

\bibitem[Kurucz(1993)]{Kurucz1993} Kurucz, R.~L.\ 1993, Kurucz CD-ROM

\bibitem[Lillo-Box et al.(2018)]{LilloBox2018} Lillo-Box, J., Barrado, D., Figueira, P., et al.\ 2018, \aap, 609, A96

\bibitem[Lillo-Box et al.(2015)]{LilloBox2015} Lillo-Box, J., Barrado, D., Santos, N.~C., et al.\ 2015, \aap, 577, A105

\bibitem[Lin et al.(1996)]{Lin1996} Lin, D.~N.~C., Bodenheimer, P., \& Richardson, D.~C.\ 1996, \nat, 380, 606


\bibitem[Mayor et al.(2003)]{Mayor2003} Mayor, M., Pepe, F., Queloz, D., et al.\ 2003, The Messenger, 114, 20

\bibitem[Mayor, \& Queloz(1995)]{MayorQueloz1995} Mayor, M., \& Queloz, D.\ 1995, \nat, 378, 355

\bibitem[McLaughlin(1924)]{McLaughlin1924} McLaughlin, D.~B.\ 1924, \apj, 60, 22

\bibitem[Millholland et al.(2016)]{Millholland2016} Millholland, S., Wang, S., \& Laughlin, G.\ 2016, \apjl, 823, L7


\bibitem[Miralda-Escud{\'e}(2002)]{MiraldaEscude2002} Miralda-Escud{\'e}, J.\ 2002, \apj, 564, 1019


\bibitem[Naoz et al.(2011)]{Naoz2011} Naoz, S., Farr, W.~M., Lithwick, Y., et al.\ 2011, \nat, 473, 187


\bibitem[{\"O}berg et al.(2011)]{Oberg2011} {\"O}berg, K.~I., Murray-Clay, R., \& Bergin, E.~A.\ 2011, \apjl, 743, L16

\bibitem[Oshagh et al.(2015)]{Oshagh2015} Oshagh, M., Santos, N.~C., Figueira, P., et al.\ 2015, \aap, 583, L1

\bibitem[Petigura et al.(2018)]{Petigura2018} Petigura, E.~A., Marcy, G.~W., Winn, J.~N., et al.\ 2018, \aj, 155, 89

\bibitem[Pont et al.(2009)]{Pont2009} Pont, F., Gilliland, R.~L., Knutson, H., et al.\ 2009, \mnras, 393, L6

\bibitem[Queloz et al.(2000)]{Queloz2000} Queloz, D., Eggenberger, A., Mayor, M., et al.\ 2000, \aap, 359, L13

\bibitem[Ribas et al.(2008)]{Ribas2008} Ribas, I., Font-Ribera, A., \& Beaulieu, J.-P.\ 2008, \apjl, 677, L59

\bibitem[Ricker et al.(2015)]{Ricker2015} Ricker, G.~R., Winn, J.~N., Vanderspek, R., et al.\ 2015, Journal of Astronomical Telescopes, Instruments, and Systems, 1, 014003

\bibitem[Rodriguez et al.(2019)]{Rodriguez2019} Rodriguez, J.~E., Quinn, S.~N., Huang, C.~X., et al.\ 2019, \aj, 157, 191

\bibitem[Rossiter(1924)]{Rossiter1924} Rossiter, R.~A.\ 1924, \apj, 60, 15

\bibitem[Rossum et al.(1995)]{Rossum1995} Rossum, G.\ 1995, Python Reference Manual

\bibitem[Schlafly \& Finkbeiner(2011)]{Schlafly2011} Schlafly, E.~F., Finkbeiner, D.~P., \ 2011, \apj, 737, 103

\bibitem[Schlegel et al.(1998)]{Schlegel1998} Schlegel, D.~J., Finkbeiner, D.~P., \& Davis, M.\ 1998, \apj, 500, 525 

\bibitem[Schlesinger(1910)]{Schlesinger1910} Schlesinger, F.\ 1910, Publications of the Allegheny Observatory of the University of Pittsburgh, 1, 123

\bibitem[Sing et al.(2016)]{Sing2016} Sing, D.~K., Fortney, J.~J., Nikolov, N., et al.\ 2016, \nat, 529, 59

\bibitem[Schuessler \& Solanki(1992)]{SchuesslerSolanki1992} Schuessler, M., \& Solanki, S.~K.\ 1992, \aap, 264, L13

\bibitem[Smalley et al.(2011)]{Smalley2011} Smalley, B., Anderson, D.~R., Collier Cameron, A., et al.\ 2011, \aap, 526, A130

\bibitem[Smith et al.(2012)]{Smith2012} Smith, J.~C., Stumpe, M.~C., Van Cleve, J.~E., et al.\ 2012, \pasp, 124, 1000

\bibitem[Sneden(1973)]{Sneden1973} Sneden, C.\ 1973, \apj, 184, 839 

\bibitem[Sousa et al.(2015)]{Sousa2015} Sousa, S.~G., Santos, N.~C., Adibekyan, V., Delgado-Mena, E., \& Israelian, G.\ 2015, \aap, 577, A67 

\bibitem[Stassun et al.(2017)]{Stassun2017} Stassun, K.~G., Collins, K.~A., \& Gaudi, B.~S.\ 2017, \aj, 153, 136 

\bibitem[Stassun et al.(2018)]{Stassun2018a} Stassun, K.~G., Corsaro, E., Pepper, J.~A., \& Gaudi, B.~S.\ 2018, \aj, 155, 22 

\bibitem[Stassun et al.(2018)]{Stassun2018b} Stassun, K.~G., Oelkers, R.~J., Pepper, J., et al.\ 2018, \aj, 156, 102

\bibitem[Stassun \& Torres(2016)]{StassunTorres2016} Stassun, K.~G., \& Torres, G.\ 2016, \aj, 152, 180 

\bibitem[Stassun \& Torres(2018)]{StassunTorres2018} Stassun, K. G. \& Torres, G. 2018 ApJ, 862, 61  

\bibitem[Stumpe et al.(2014)]{Stumpe2014} Stumpe, M.~C., Smith, J.~C., Catanzarite, J.~H., et al.\ 2014, \pasp, 126, 100


\bibitem[Tokovinin et al.(2013)]{Tokovinin2013} Tokovinin, A., Fischer, D.~A., Bonati, M., et al.\ 2013, \pasp, 125, 1336

\bibitem[Tokovinin(2018)]{Tokovinin2018} Tokovinin, A.\ 2018, \pasp, 130, 035002

\bibitem[Torres et al.(2004)]{Torres2004} Torres, G., Konacki, M., Sasselov, D.~D., et al.\ 2004, \apj, 614, 979

\bibitem[Torres et al.(2010)]{Torres2010} Torres, G., Andersen, J., \& Gim{\'e}nez, A.\ 2010, \aapr, 18, 67 

\bibitem[Twicken et al.(2018)]{Twicken2018} Twicken, J.~D., Catanzarite, J.~H., Clarke, B.~D., et al.\ 2018, \pasp, 130, 064502 



\bibitem[van der Walt et al.(2011)]{VanderWalt2011} van der Walt, S., Colbert, S.~C., Varoquaux, G.\ 2011, Computing in Science \& Engineering, 13, 22

\bibitem[Vogt et al.(1994)]{Vogt1994} Vogt, S.~S., Allen, S.~L., Bigelow, B.~C., et al.\ 1994, \procspie, 362

\bibitem[Wang et al.(2018)]{Wang2018} Wang, S., Addison, B., Fischer, D.~A., et al.\ 2018, \aj, 155, 70

\bibitem[Wang et al.(2019)]{Wang2019} Wang, S., Jones, M., Shporer, A., et al.\ 2019, \aj, 157, 51

\bibitem[Winn(2009)]{Winn2009} Winn, J.~N.\ 2009, Transiting Planets, 99


\bibitem[Winn \& Fabrycky(2015)]{WinnFabrycky2015} Winn, J.~N., \& Fabrycky, D.~C.\ 2015, \araa, 53, 409

\bibitem[Wu \& Lithwick(2011)]{WuLithwick2011} Wu, Y., \& Lithwick, Y.\ 2011, \apj, 735, 109

\bibitem[Wu et al.(2018)]{Wu2018} Wu, D.-H., Wang, S., Zhou, J.-L., et al.\ 2018, \aj, 156, 96



\bibitem[Zhou et al.(2019)]{Zhou2019} Zhou, G., Huang, C., Bakos, G., et al.\ 2019, arXiv e-prints, arXiv:1906.00462

\bibitem[Ziegler et al.(2019)]{Ziegler2019} Ziegler, C., Tokovinin, A., Briceno, C., et al.\ 2019, arXiv e-prints, arXiv:1908.10871

\bibitem[Zucker, \& Mazeh(1994)]{ZuckerMazeh1994} Zucker, S., \& Mazeh, T.\ 1994, \apj, 420, 806

\end{thebibliography}
\end{document}